\documentclass[twocolumn]{svjour3}
\usepackage{graphicx}
\usepackage{caption}
\usepackage{amsmath,amssymb}
\usepackage[misc,geometry]{ifsym}
\usepackage{verbatim}
\usepackage[numbers]{natbib}
\usepackage[normalem]{ulem}
\usepackage{booktabs}
\usepackage[hang,flushmargin]{footmisc}
\usepackage[english]{babel}

\def\jcp{J. Chem.\ Phys.\ }

\def\molp{Mol.\ Phys.\ }

\def\oc{Opt.\ Commun.\ }

\def\pra{Phys.\ Rev.\ A }

\def\prd{Phys.\ Rev.\ D }

\def\prl{Phys.\ Rev.\ Lett.\ }

\def\rsi{Rev.\ Sci.\ Instr.\ }

\def\jms{J. Mol.\ Spectrosc.\ }

\def\Hmol{H$_2$}
\def\HSulf{H$_2$S}
\def\Hmolx{H$_2^*$}
\def\Hmolp{H$_2^+$}
\def\Hp{H$^+$}
\def\wn{cm$^{-1}$}
\begin{document}

\title{Precision measurements and test of molecular theory \\in highly-excited vibrational states of H$_2$ $(v=11)$}

\author{T.~Madhu~Trivikram$^1$ \and M.~L.~Niu$^1$ \and P.~Wcis{\l}o$^2$ \and W.~Ubachs$^1$ \and E.~J.~Salumbides$^{1,3}$}

\institute{\Letter E. J. Salumbides\\\email{e.j.salumbides@vu.nl}\\
$^1$ Department of Physics and Astronomy, LaserLaB,\\ Vrije Universiteit Amsterdam, De Boelelaan 1081,\\ 1081 HV Amsterdam, The Netherlands\\
$^2$ Institute of  Physics,  Faculty  of  Physics, Astronomy and Informatics, Nicolaus Copernicus University,\\ Grudzi\c{a}dzka 5, PL-87-100 Toru\'n, Poland\\
$^3$ Department of Physics, University of San Carlos,\\ Cebu City 6000, Philippines
}

\date{\today}

\titlerunning{Precision measurements and test of molecular theory in H$_2$ ($v=11$)}
\maketitle

\begin{abstract}

Accurate $EF{}^1\Sigma^+_g-X{}^1\Sigma^+_g$ transition energies in molecular hydrogen were determined for transitions originating from levels with highly-excited vibrational quantum number, $v=11$, in the ground electronic state.
Doppler-free two-photon spectroscopy was applied on vibrationally excited H$_2^*$, produced via the photodissociation of H$_2$S, yielding transition frequencies with accuracies of $45$ MHz or $0.0015$ cm$^{-1}$.
An important improvement is the enhanced detection efficiency by resonant excitation to autoionizing $7p\pi$ electronic Rydberg states, resulting in narrow transitions due to reduced ac-Stark effects.
Using known $EF$ level energies, the level energies of $X(v=11, J=1,3-5)$ states are derived with accuracies of typically 0.002 cm$^{-1}$.
These experimental values are in excellent agreement with, and are more accurate than the results obtained from the most advanced \emph{ab initio} molecular theory calculations including relativistic and QED contributions.

\end{abstract}

\section{Introduction}

The advance of precision laser spectroscopy of atomic and molecular systems has, over the past decades, been closely connected to the development of experimental techniques such as tunable laser technology~\cite{Haensch1972}, saturation spectroscopy~\cite{Haensch1971}, two-photon Doppler-free spectroscopy~\cite{Haensch1975a}, cavity-locking techniques~\cite{Haensch1980}, and ultimately, the invention of the frequency comb laser~\cite{Holzwarth2000}, developments to which Prof. Theodore H\"{a}nsch has greatly contributed.
These inventions are being exploited to further investigate at ever-increasing precision the benchmark atomic system -- the hydrogen atom, evinced by the advance in spectroscopic accuracy of atomic hydrogen measurements by more than seven orders of magnitude since the invention of the laser \cite{Haensch2006}.
The spectroscopy of the 1S-2S transition in atomic hydrogen, at $4\times10^{-15}$ relative accuracy \cite{Parthey2011}, provides a stringent test of fundamental physical theories, in particular quantum electrodynamics (QED).
Currently, the theoretical comparison to precision measurements on atomic hydrogen are limited by uncertainties in the proton charge radius $r_p$. The finding that the $r_p$-value obtained from muonic hydrogen spectroscopy is in disagreement by some 7-$\sigma$ \cite{Antognini2013} is now commonly referred to as the \emph{proton-size puzzle}.

Molecular hydrogen, both the neutral and ionic varieties, are benchmark systems in molecular physics, in analogy to its atomic counterpart.
Present developments in the \emph{ab initio} theory of the two-electron neutral \Hmol\ molecule and the one-electron ionic \Hmolp\ molecule, as well as the respective isotopologues, have advanced in accuracy approaching that of its atomic counterpart despite the increased complexity.
The most accurate level energies of the entire set of rotational and vibrational states in the ground electronic state of \Hmol\ were calculated by Komasa et al.~\cite{Komasa2011}.
An important breakthrough in these theoretical studies was the inclusion of higher-order relativistic and QED contributions, along with a systematic assessment of the uncertainties in the calculation.
Recently, further improved calculations of the adiabatic~\cite{Pachucki2014} as well as non-adiabatic~\cite{Pachucki2015} corrections have been performed, marking the steady progress in this field.

In the same spirit as in atomic hydrogen spectroscopy, the high-resolution experimental investigations in molecular hydrogen are aimed towards confronting the most accurate \emph{ab initio} molecular theory.
For the \Hmolp\ and HD$^+$ ions, extensive efforts by Korobov and co-workers over the years, have recently led to the theoretical determination of ground electronic state level energies at 0.1 ppb accuracies \cite{Korobov2014}.
The latter accuracy enables the extraction of the proton-electron mass ratio, $m_p/m_e$, when combined with the recent HD$^+$ spectroscopy using a laser-cooled ion trap \cite{Biesheuvel2016}. These are currently at lower precision than other methods but prospects exist that competitive values can be derived from molecular spectroscopy.
Similarly Karr et al.~\cite{Karr2016} recently discussed the possibility of determining $R_\infty$ using \Hmolp\ (or HD$^+$) transitions as an alternative to atomic hydrogen spectroscopy.
Even for the neutral system of molecular hydrogen, the determination of $r_p$ from spectroscopy is projected to be achievable, from the ongoing efforts in both calculation~\cite{Pachucki2016} and experiments~\cite{Ubachs2016}.
Molecular spectroscopy might thus be posed to contribute towards the resolution of the proton-size puzzle.

In contrast to atomic structure, the added molecular complexity due to the vibrational and rotational nuclear degrees of freedom could constitute an important feature, with a multitude of transitions (in the ground electronic state) that can be conscripted towards the confrontation of theory and experiments.
From both experimental and theoretical perspectives, this multiplicity allows for consistency checks and assessment of systematic effects.
In recent years, we have tested the most accurate \Hmol\ quantum chemical calculations using various transitions, for example, the dissociation limit $D_0$ or binding energy of the ground electronic state \cite{Liu2009}; the rotational sequence in the vibrational ground state \cite{Salumbides2011}; and the determination of the ground tone frequency $(v=0\rightarrow 1)$ \cite{Dickenson2013}.
The comparisons exhibit excellent agreement thus far, and have in turn been interpreted to provide constraints of new physics, such as fifth-forces \cite{Salumbides2013} or extra dimensions \cite{Salumbides2015b}.

Recently, we reported a precision measurement on highly-excited vibrational states in \Hmol\ \cite{Niu2015b}.
The experimental investigation of such highly-excited vibrational states probe the region where the calculations of Komasa, Pachucki and co-workers~\cite{Komasa2011} are the least accurate, specifically in the $v=6-12$ range.
The production of excited \Hmolx\ offers a unique possibility on populating the high-lying vibrational states, that would otherwise be practically inaccessible by thermodynamic means (corresponding temperature of $T\sim 47,000$ K for $v=11$).

Here, we present measurements of level energies of $v=11$ rovibrational quantum states that extend the spectroscopy in Ref.~\cite{Niu2015b} and that implement improvements, leading to a narrowing of the resonances.
This is achieved by the use of a resonant ionization step to molecular Rydberg states, thereby enhancing the detection efficiency significantly.
The enhancement allows for the use of a low intensity spectroscopy laser minimizing the effect of ac-Stark induced broadening and shifting of lines.
The ac-Stark effect is identified as the major source of systematic uncertainty in the measurements, and a detailed treatment of this phenomenon is also included in this contribution.

\section{Experiment}

\begin{figure}[!b]
\centering
\includegraphics[width=0.5\textwidth]{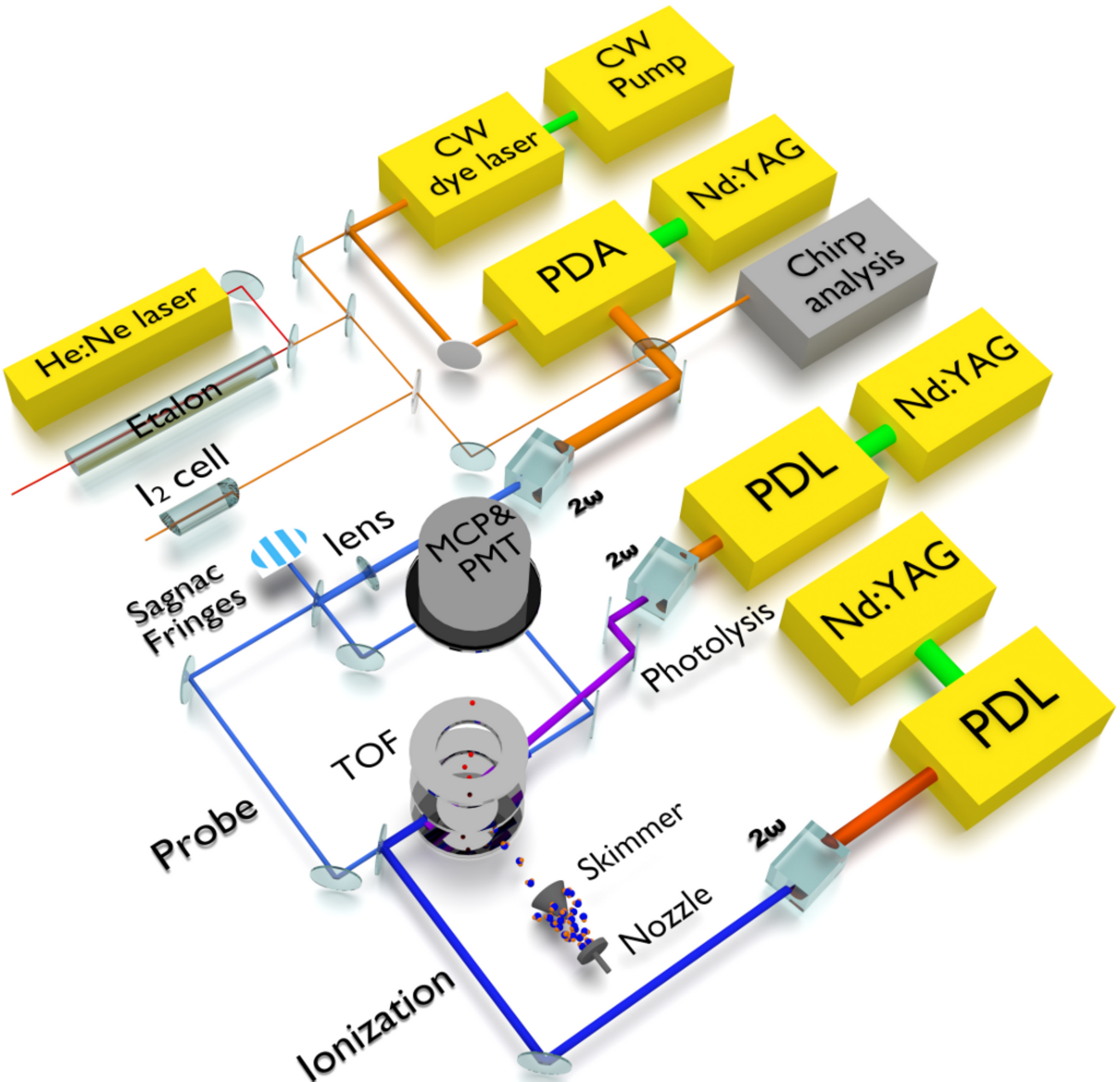}
\caption{Schematic of experimental setup indicating the three main radiation sources: the \emph{photolysis} laser at 293 nm for H$_2^*$ production from H$_2$S dissociation; the \emph{probe} laser from a narrowband dye amplifier (PDA) for the $EF-X$ spectroscopy transition; and the \emph{ionization} laser for resonant H$_2^+$ ion production, subsequently detected as signal.
Doppler-free two-photon excitation is facilitated by the Sagnac interferometric alignment of the counter-propagating probe beams.
Absolute frequency calibration is performed with respect to I$_2$ hyperfine reference lines, aided by the relative frequency markers from the transmission fringes of a length-stabilized Fabry-P\'{e}rot etalon.
The frequency offset between the cw-seed and PDA pulse output, induced by chirp effects in the dye amplifier, is measured and corrected for, post-measurement.
[cw: continuous wave; PDL: pulsed dye laser; $2\omega$: frequency-doubling stage; TOF: time-of-flight region; MCP: multichannel plates; PMT: photomultiplier tube]
}
\label{fig:setup}
\end{figure}

The production of excited \Hmolx\ from the photodissociation of hydrogen sulfide was first demonstrated by Steadman and Baer~\cite{Steadman1989}, who observed that the nascent \Hmolx\ molecules were populated at predominantly high vibrational quanta in the two-photon dissociation of \HSulf\ at UV wavelengths.  That study used a single powerful laser for dissociation, for subsequent \Hmol\ spectroscopy, and to induce dissociative ionization for signal detection.
Niu et al.~\cite{Niu2015b} utilized up to three separate laser sources to address the production, probe, and detection steps in a better controlled fashion. The present study, targeting \Hmol$(v=11)$ levels, is performed using the same experimental setup as in Ref.~\cite{Niu2015b}, depicted schematically in Fig.~\ref{fig:setup}. The photolysis laser at $293$ nm, generated from the second harmonic of the output of a commercial pulsed dye laser (PDL) with Rhodamine B dye, serves in the production of \Hmolx\ by photolysing \HSulf. The narrowband spectroscopy laser radiation at around $300-304$ nm is generated by frequency upconversion of the output of a continuous wave (cw)-seeded pulsed dye amplifier system (PDA) running on Rhodamine 640 dye. The ionization laser source at $302-305$ nm, from the frequency-doubled output of another PDL (also with Rhodamine 640 dye), is used to resonantly excite from the $EF$ to autoionizing Rydberg states to eventually form \Hmolp\ ions. This $2+1'$ resonance-enhanced multiphoton ionization (REMPI) scheme results in much improved sensitivities compared to our previous study~\cite{Niu2015b}.

The \HSulf\ molecular beam, produced by a pulsed solenoid valve in a source vacuum chamber, passes through a skimmer towards a differentially pumped interaction chamber, where it intersects the laser beams perpendicularly. The probe or spectroscopy laser beam is split into two equidistant paths and subsequently steered in a counter-propagating orientation, making use of a Sagnac interferometer alignment for near-perfect cancellation of the residual first-order Doppler shifts~\cite{Hannemann2007}. Moreover, the probe laser beams pass through respective lenses, of $f=50$-cm focal length, to focus and enhance the probe intensity at the interaction volume. Finally, the ionization beam is aligned in almost co-linear fashion with the other laser beams to ensure maximum spatial overlap.
To avoid ac-Stark shifts during the spectroscopic interrogation, induced by the photolysis laser ($\sim6$ mJ typical pulse energy; $\sim10$-ns pulse duration), a 15-ns delay between the photolysis and probe pulses is established with a delay line.
For a similar reason, the ionization pulse ($\sim1$ mJ typical pulse energy; $\sim10$-ns pulse duration) is also delayed by 30 ns with respect to the probe pulse.
The 1~mJ ionization pulse energy is sufficient for saturating the ionization step.

The ions produced in the interaction volume are accelerated by ion lenses, further propagating through a field-free time-of-flight (TOF) mass separation region before impinging on a multichannel plate (MCP) detection system. Scintillations in a phosphor screen behind the MCP are monitored by a photomultiplier tube (PMT) and a camera, culminating in the recording of the mass-resolved signals.
In the non-resonant ionization step as in Ref.~\cite{Niu2015b}, predominantly \Hp\ ions were produced and were thus used as the signal channel for the $EF-X$ excitation.
In contrast, the resonant ionization scheme employed here predominantly produces \Hmolp\ ions. In addition to the enhancement of sensitivity, the \Hmolp\ channel offers another important advantage as it is a background free channel, whereas the \Hp\ channel includes significant contributions from \HSulf, as well as SH, dissociative ionization products.
To avoid dc-Stark effects on the transition frequencies, the acceleration voltages of the ion lens system are pulsed and time-delayed with respect to the probe laser excitation.

Niu et al.~\cite{Niu2015b} confirmed the observation of \Hmol\ two-photon transitions in various $EF-X$ ($v'$,10-12) bands, first identified by Steadman and Baer~\cite{Steadman1989}, but only for transitions to the outer $F-$well of the $EF$ electronic potential in H$_2$.
Franck-Condon factor (FCF) calculations, to assess the transition strengths of the photolysis-prepared levels of $X(v'')$ to levels in the combined inner ($E$) and outer ($F$) wells of the $EF$ double well potential, were performed by Fantz and
W\"underlich \cite{Fantz2006,Fantz2004}. 
While Niu et al.~\cite{Niu2015b} performed precision measurements probing the $X(v''=12)$ levels, presently $X(v''=11)$ levels are probed. Note that for the excited state two different numberings of vibrational levels exist: one counting the levels in the combined $EF$ well, and the other counting the levels in the $E$ and $F$ wells separately.
Then $EF(v'=5)$ corresponds to $F(v'=3)$, while $EF(v'=1)$ corresponds to $F(v'=0)$.
In the following, we will refer to the vibrational assignments following the $F$-well notation.
The FCF for the $F-X (3,12)$ band, used in Ref.~\cite{Niu2015b}, amounts to 0.047~\cite{Fantz2004} and that of the presently used $F-X (0,11)$ band amounts to 0.17, making the latter band's transitions three times stronger.

\begin{figure}[!b]
\centering
\includegraphics[width=0.45\textwidth]{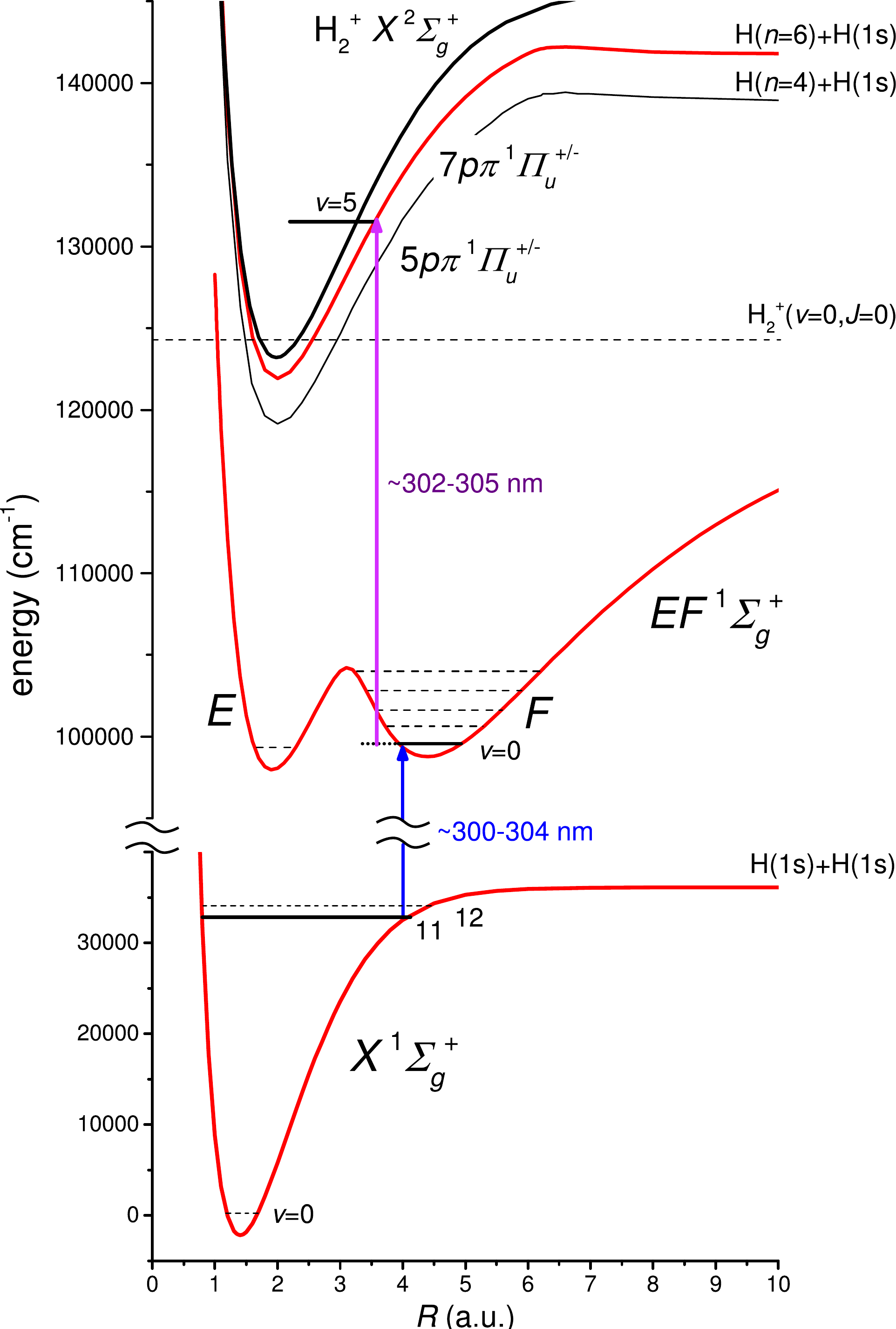}
\caption{
Potential energy diagram showing the relevant \Hmol\ electronic states in the $2+1'$ REMPI study.
Two-photon Doppler-free spectroscopy is applied on the $F-X$ (0,11) band.
Resonant excitation to the $7p\pi$ state by the detection laser follows the spectroscopic excitation, leading to subsequent autoionization yielding enhanced \Hmolp\ ion signal.
}
\label{fig:potential}
\end{figure}

\subsection{Resonant ionization}

The non-resonant ionization step was the major limitation in \cite{Niu2015b}, since this prohibits the spectroscopy to be carried out at sufficiently low probe laser intensities.
Due to ac-Stark effects, the lines were broadened to more than 1 GHz, while the expected instrumental linewidth is less than 200 MHz.
Moreover, at higher probe intensities asymmetric line profiles are observed, reducing the accuracy of the line position determination and ultimately limiting the ac-Stark extrapolation to the unperturbed line position. 

While signal improvement was observed when employing a detection laser in the range between 202-206 nm in Ref.~\cite{Niu2015b}, the enhancement was limited since no sharp resonances were found, indicating excitation to some continuum.
For the present study, a thorough search for resonances from the $F$ state was undertaken. The $np\pi$ and $np\sigma$ Rydberg series, with principal quantum number $n=5-7$ were identified as potential candidates based on the FCFs for the outer $F$-well. The search was based on reported FCFs for the $D^1\Pi_u - F^1\Sigma_g^+$ ($v'$,1) bands~\cite{Fantz2004}.
It was further assumed that the FCFs for the $np\pi\,^1\Pi_u - F^1\Sigma_g^+$ electronic systems are comparable to that of $3p\pi\,D^1\Pi_u - F^1\Sigma_g^+$, since the potential energy curves for the $np\pi\,^1\Pi_u$ Rydberg states are similar as they all converge to the \Hmolp\ ionic potential.
Note the particular characteristic of the $n=5-7$ $np\,^1\Pi_u$ Rydberg states, that dissociate to a ground state atom and another with a principal quantum number H($n-1$),
i.e. $5p\pi\rightarrow$ H(1s) + H(4f); $6p\pi\rightarrow$ H(1s) + H(5f); $7p\pi\rightarrow$ H(1s) + H(6d) \cite{Glass-Maujean2013b,Glass-Maujean2013a,Glass-Maujean2013c}.
The electron configuration changes as a function of the internuclear distance $R$, e.g. the low vibrational levels of the $5p\,^1\Pi_u$ follow a diabatic potential that extrapolates to the $n=5$ limit, and not the $n=4$ dissociation limit at $R\rightarrow\infty$.   
This peculiarity is explained by $l$-uncoupling \cite{Herzberg1972}, as the molecule changes from Hund's case b (Born-Oppenheimer) to Hund's case d (complete nonadiabatic mixing) with increasing principal quantum number, corresponding to the independent nuclear motion of the residual ion core \Hmolp\ and the excited Rydberg electron $np$. 
Such transition in Hund's cases occurs at $n\sim 7$ for low rotational quantum numbers \cite{Glass-Maujean2013a}.

\begin{figure}[!t]
\includegraphics[width=0.48\textwidth]{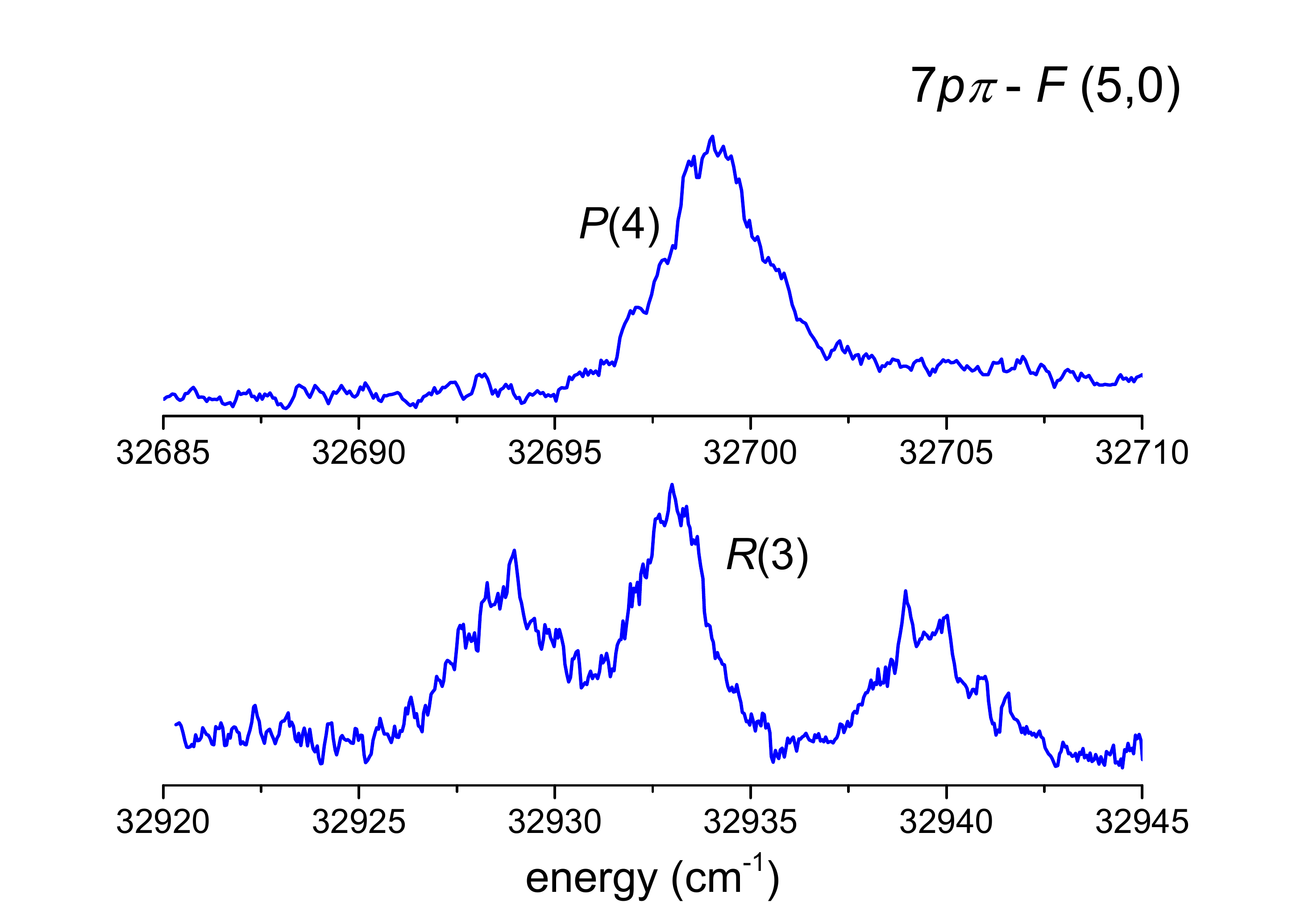}
\caption{Autoionizing $R(3)$ and $P(4)$ resonances of the $7p\pi\,^1\Pi_u - F\,^1\Sigma_g^+$ (5,0) band which are crucial in enhancing the \Hmolp\ signal strength of the $F-X$ spectroscopy .}
\label{fig:rydberg}
\end{figure}

The $np\pi\,^1\Pi_u$ Rydberg states decay via three competing channels: by fluorescence to lower $n$ electronic configurations; by predissociation, where the nascent H atom is further photoionized to yield \Hp\ ions; and lastly, by autoionization to yield \Hmolp\ ions. Recent analysis of one-photon absorption measurements using XUV synchrotron radiation in the range of $74-81$ nm demonstrated that autoionization completely dominates over the other two competing channels in the case of $n>5$~\cite{Glass-Maujean2010a}. Using the $n=5-7$ level energies of the Rydberg levels reported in \cite{Glass-Maujean2013b,Glass-Maujean2013a,Glass-Maujean2013c}, the detection laser was scanned in the vicinity of the expected transition energies, where it turned out that transitions to $7p\pi,\, v=5$ resulted in sufficient \Hmolp\ signal enhancement.
The maximum FCF overlap for the $D^1\Pi_u - F^1\Sigma_g^+$ system is $0.34$ for the (8,0) band, while the corresponding $D-F$ (5,0) band only has an FCF of $0.022$ ~\cite{Fantz2004}.
Although the (8,0) band with better FCF could be used, the ionization step is already saturated using the weaker (5,0) band.
Autoionization resonances are shown in Fig.~\ref{fig:rydberg} and assigned to $R(3)$ and $P(4)$ lines in the $7p\pi\,^1\Pi_u - F\,^1\Sigma_g^+$ (5,0) bands, whose widths are in good agreement with the synchrotron data~\cite{Glass-Maujean2013b}.
The neighboring resonances of the $R(3)$ line in Fig.~\ref{fig:rydberg} are not yet assigned but is not relevant to the $F-X$ investigation here.
Appropriate $7p\pi - F$ transitions are used for the ionization of particular $F-X$ two-photon $Q(J)$ transitions.

\subsection{Frequency calibration}

\begin{figure}[!t]
\includegraphics[width=0.5\textwidth]{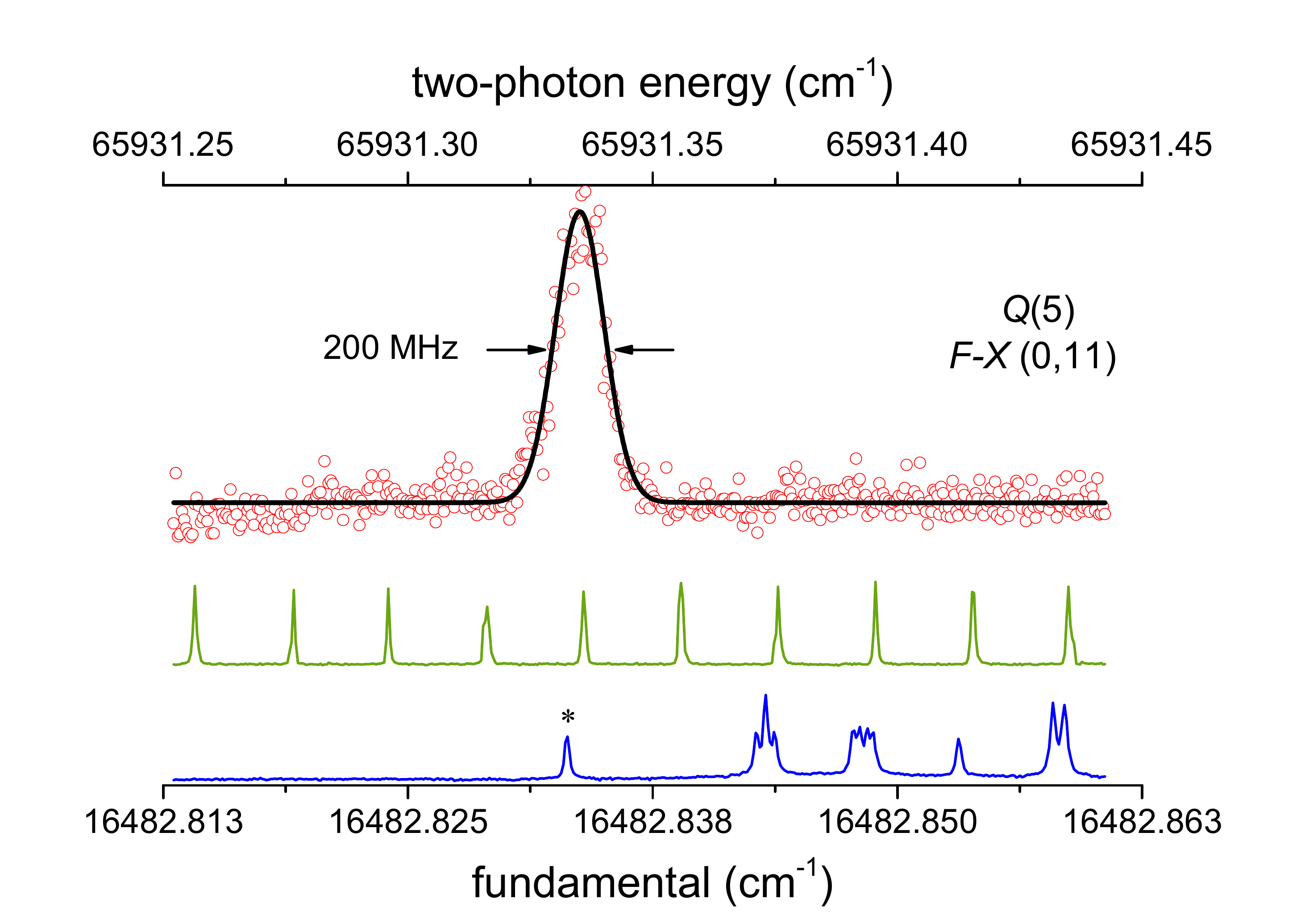}
\caption{Recording of the $F\,^1\Sigma_g^+ - X\,^1\Sigma_g^+$ (0,11) $Q(5)$ transition is shown at a probe laser intensity of 280 MW/cm$^2$ and detection in the \Hmolp\ signal channel.
The transmission markers of a length-stabilized Fabry-P\'{e}rot etalon (FSR=148.96 MHz) are used in the relative frequency calibration, while the hyperfine feature marked with * of the I$_2$ $B-X(11,2)\,P(94)$ transition serves as an absolute frequency reference (* at 16482.83312 cm$^{-1}$).
}
\label{fig:hires_spectra}
\end{figure}

A representative high-resolution spectrum of the $F\,^1\Sigma_g^+$--$X\,^1\Sigma_g^+$ (0,11) $Q(5)$ transition taken at low probe intensity is displayed in Fig.~\ref{fig:hires_spectra}.
Simultaneous with the \Hmol\ spectroscopy, the transmission fringes of the PDA cw-seed radiation through a Fabry-P\'{e}rot interferometer were also recorded to serve as relative frequency markers, with the free spectral range of FSR=148.96(1) MHz. The etalon is temperature-stabilized and its length is actively locked to a frequency-stabilized HeNe laser. The absolute frequency calibration is obtained from the I$_2$ hyperfine-resolved saturation spectra using part of the cw-seed radiation. For the $Q(5)$ line in Fig.~\ref{fig:hires_spectra}, the I$_2$ $B-X(11,2)\,P(94)$ transition is used, where the line position of the hyperfine feature marked with an * is $16\,482.833\,12(1)$ \wn\ \cite{Xu2000, Bodermann2002}. The accuracy of the frequency calibration for the narrow \Hmol\ transitions is estimated to be 1 MHz in the fundamental or 4 MHz in the transition frequency, after accounting for a factor of 4 for the harmonic up-conversion and two-photon excitation.

For sufficiently strong transitions probed at the lowest laser intensities, linewidths as narrow as 150 MHz were obtained.
This approaches the Fourier-transform limited instrumental bandwidth of 110 MHz, for the 8-ns pulsewidths at the fundamental, approximated to be Gaussian, that also includes a factor two to account for the frequency upconversion. 
The narrow linewidth obtained demonstrates that despite of the photodissociation process imparting considerable kinetic energy on the produced \Hmolx, additional Doppler-broadening is not observed.
Although not unexpected due to the Doppler-free experimental scheme implemented, this strengthens the claim that residual Doppler shifts are negligible.

\begin{figure}[!t]
\includegraphics[width=0.45\textwidth]{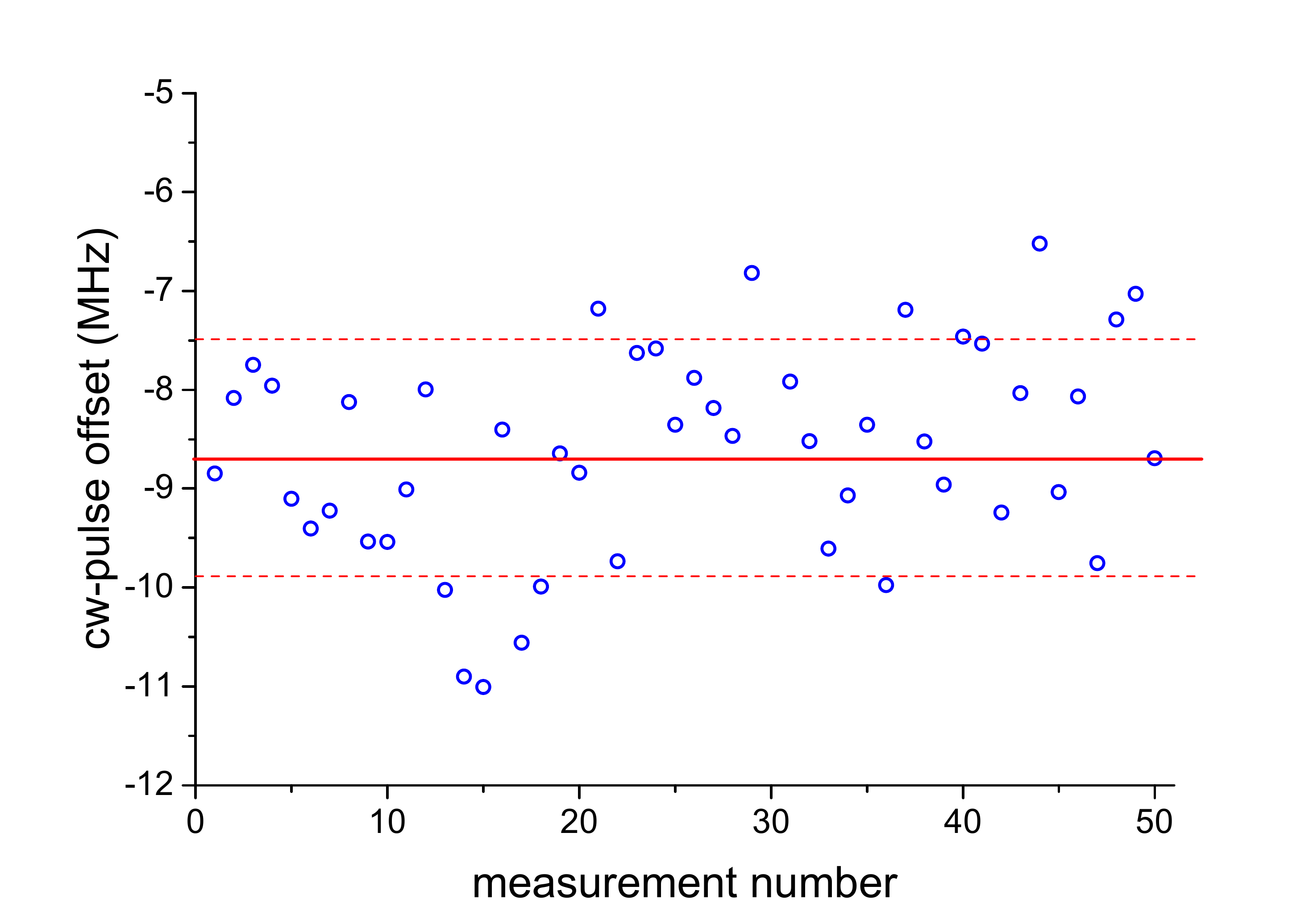}
\caption{Chirp-induced cw--pulse frequency offset of the fundamental radiation for a fixed PDA wavelength. The solid line indicates the average and the dashed lines indicate the standard deviation.}
\label{fig:chirp}
\end{figure}

Since the frequency calibration is performed using the cw-seed while the spectroscopy is performed using the PDA output pulses, any cw--pulse frequency offset need to be measured and corrected for~\cite{Niu2015}.
A typical recording of the chirp-induced frequency offset for a fixed PDA wavelength is shown in Fig.~\ref{fig:chirp}. While the measurements can be done online for each pulse, this comes at the expense of a slower data acquisition speed, and was only implemented for a few recordings in order to assess any systematic effects.
A flat profile of the cw-pulse offset when the wavelength was tuned over the measurement range accessed in this study justifies this offline correction.
Typical cw--pulse frequency offset values were measured to be $-8.7(1.2)$ MHz in the fundamental, which translates to $-35(5)$ MHz 
in the transition frequency.

\subsection{Uncertainty estimates}

The sources of uncertainties and the respective contributions are shown in Table~\ref{tab:uncertainty}. The contributions of each source are summed in quadrature in order to obtain the final uncertainty for each transition.
Data sets from which separate ac-Stark extrapolations to zero-power were performed on different days, and were verified to exhibit consistency within the statistical uncertainty of 0.0014 \wn.
Note that the estimates shown in Table~\ref{tab:uncertainty} are only for the low probe intensity measurements used to obtain the highest resolutions.
The uncertainties of the present investigation constitute more than a factor of two improvement over our previous study in \cite{Niu2015b}.
The dominant source of systematic uncertainty is the ac-Stark shift and is discussed in more detail in the following section.

\section{ac-Stark shift and broadening}

In the perturbative regime, the leading-order energy level shift $\Delta E_n$ of a state $|n\rangle$ induced by a linearly-polarized optical field with an amplitude $\mathcal{E}_0$, and frequency $\nu$ can be described as
\begin{align}
\Delta E_n=\frac{1}{2} \sum_m &\left\{\frac{\langle n|\overrightarrow{\mu}\cdot\overrightarrow{\mathcal{E}_0}|m\rangle\langle m|\overrightarrow{\mu}\cdot\overrightarrow{\mathcal{E}_0}|n\rangle}{E_n - E_m -h\nu} \right. \nonumber\\
     &\left. + \frac{\langle n|\overrightarrow{\mu}\cdot\overrightarrow{\mathcal{E}_0}|m\rangle\langle m|\overrightarrow{\mu}\cdot\overrightarrow{\mathcal{E}_0}|n\rangle}{E_n - E_m + h\nu} \right\},
\label{eq:deltaE}
\end{align}
where $\langle n|\mu|m\rangle$ is the transition dipole moment matrix element between states $n$ and $m$, with an energy $E_m$ for the latter~\cite{Bakos1977}.
Thus $\Delta E_n$ has a quadratic dependence on the field, or a linear dependence on intensity for this frequency-dependent ac-Stark level shift.
In a simple case, when there is one near-resonant coupling to state $m$ whose contribution dominates $\Delta E_n$, the sign of the detuning with respect to transition frequency $\nu_{mn}=|E_n-E_m|/h$ determines the direction of the light shifts with intensity.
When the probing radiation is blue-detuned, i.e. $\nu > \nu_{mn}$, the two levels $|n\rangle$ and $|m\rangle$ shift toward each other, while for red-detuning, $\nu < \nu_{mn}$, the levels repel each other.
In the case when all accessible states are far off-resonant, both terms in Eq.~(\ref{eq:deltaE}) contribute for each state $m$, and numerous $m$-states need to be included in the calculations to explain the magnitude and sign of $\Delta E_n$.
The energy shifts of the upper ($\Delta E_u$) and lower ($\Delta E_l$) levels in turn translate into an ac-Stark shift,
\begin{align}
h\delta_S=\alpha I,
\label{eq:Stark_coeff}
\end{align}
where $\alpha$ is the ac-Stark coefficient.
The measured \emph{transition} energy is $h\nu = h\nu_0 + \alpha I$, where $\nu_0=|E_u-E_l|/h$ is the unperturbed (zero-field) transition frequency.
The ac-Stark coefficient $\alpha$ depends on the coupling strengths of the $|u\rangle$ and $|l\rangle$ levels to the dipole-accessible $|m\rangle$ states, as well as the magnitude and sign of the detuning. 
We note that in a so-called \emph{magic wavelength} configuration, the frequency $\nu$ is selected so that the level shifts of the upper and lower states cancel out, leading to ac-Stark free transition frequencies \cite{Ye2008}. 

\begin{table}
\caption{Uncertainty contributions in units of $10^{-3}$ \wn.}
\label{tab:uncertainty}
\begin{minipage}{\linewidth}
\renewcommand\footnoterule{}
\begin{tabular}{lcc}
\toprule
Source			&Correction	&Uncertainty \\ \midrule
line-fitting		&--		& 0.5 \\
ac-Stark\footnote{correction depends on transition} 		&--& 1.0 \\
frequency calibration	&--		& 0.3 \\
cw--pulse offset 	&-1.2		& 0.2  \\
residual Doppler 	&0		& $<0.1$ \\
dc-Stark 		&0		& $<0.1$ \\ \midrule
total			& 		& 1.5\\ \bottomrule
\end{tabular}
\end{minipage}
\end{table}

The first experimental study of ac-Stark effects in molecules associated with REMPI processes was performed by Otis and Johnson on NO \cite{Otis1981}.
The broad ac-Stark-induced features in NO were later explained in the extensive models by Huo et al. \cite{Huo1985}.
An investigation of ac-Stark effects on two-photon transitions in CO was performed by Girard et al. \cite{Girard1983}. 
For molecular hydrogen, various studies have been performed on the two-photon excitation in the $EF-X$ system over the years \cite{Dickenson2013,Srinivasan1983b,Vrakking1993,Yiannopoulou2006,Hannemann2006}.
In the following, we present our evaluation of the ac-Stark effect in $F-X$ (0,11) transitions where we first discuss line shape effects.
This is followed by a discussion on the ac-Stark coefficients extracted from the analysis and comparisons with previous determinations on the $EF-X$ system.

\subsection{Line shape model}

\begin{figure}[!t]
\includegraphics[width=0.5\textwidth]{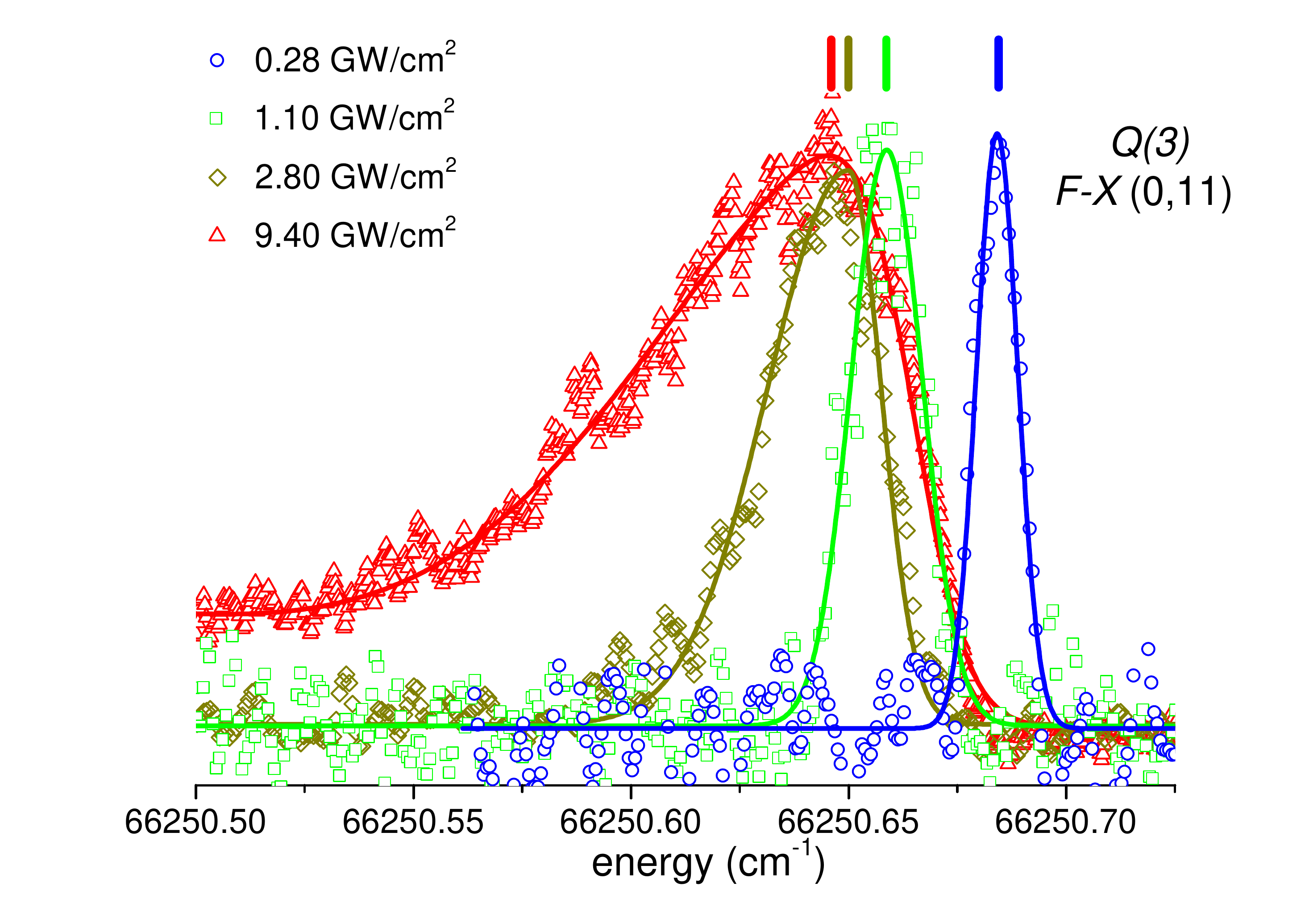}
\caption{Line profiles of the $F-X\,(0,11)\,Q(3)$ transition recorded at different probe laser intensities.
The vertical lines above each profile denote the peak position used in the subsequent extrapolation.
}
\label{fig:line_profiles}
\end{figure}

The line profiles of the $F-X$ (0,11) $Q(3)$ transition, recorded at different probe laser intensities, are displayed in Fig.~\ref{fig:line_profiles}.
The ac-Stark broadening and asymmetry is readily apparent at higher intensities, and only at low intensities can the profile be fitted by a simple Gaussian line shape.
Note that the shift in peak position at the highest probe intensity amounts to several linewidths of the lowest-intensity recording.

The asymmetry at high intensities is highly problematic with regards to the extraction of the line positions.
In our previous study~\cite{Niu2015b}, a skewed Gaussian function $g(f)$ was used to fit the spectra,
\begin{align}
g(f)=&\frac{A}{\Gamma_{G}}\exp{\left(\frac{-(f-f_c)^2}{2\Gamma^2_{G}}\right)} \nonumber\\
	&\quad\times\left\{1+\mathrm{erf}\left(\xi\,\frac{f-f_c}{\sqrt{2}\Gamma_{G}}\right)\right\},
\label{eq:skewed}
\end{align}
where $f_c$ is the Gaussian peak position in the absence of asymmetry, $\Gamma_{G}$ is the linewidth, $A$ is an amplitude scaling parameter and $\xi$ is the asymmetry parameter.
The center $f_c$ of the error function, $\mathrm{erf}(f)$, is arbitrarily chosen to coincide with the Gaussian center.
For sufficiently high intensities, a satisfactory fit is only possible if a linear background $B(f)$ is added, and thus a revised fitting function, $g'(f)=g(f)+B(f)$, is used.
This phenomenological fit function resulted in better fits than symmetric Lorentzian, Gaussian or Voigt profiles.
However, since it does not include any consideration of the underlying physics, the interpretation of the extracted $f_c$ and $\xi$ parameters, as well as the background $B(f)$, is not straightforward.
The energy position of the skewed profile maximum, instead of $f_c$, is used as the ac-Stark shifted frequency in the subsequent extrapolations.

\begin{figure}[!t]
\includegraphics[width=0.5\textwidth]{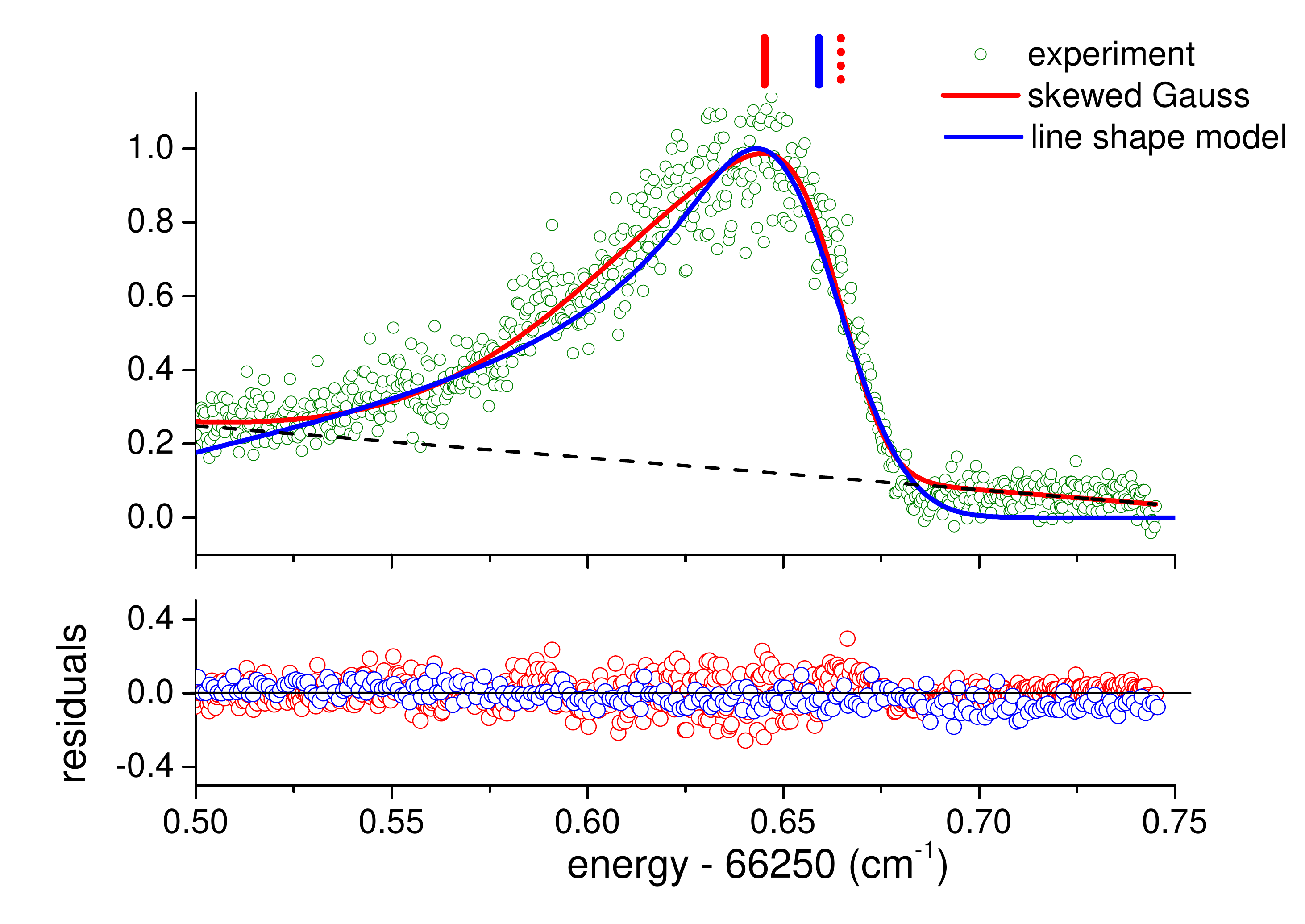}
\caption{$F-X\,(0,11)\,Q(3)$ transition taken at high probe intensity ($9.4$ GW/cm$^2$) fitted with a skewed Gaussian (red solid line), with the fitted linear background indicated by the black dashed line.
The fitted curve for line shape model is plotted as the blue solid line, which includes the effects of spatial and temporal intensity distribution.
The red vertical line above the profile indicates the line position for the skewed Gaussian fit, while the blue vertical line indicates position shifted by $\delta_0$ obtained from the line shape model.
For comparison, the $f_c$ parameter obtained from skewed Gaussian fitting is also indicated (dotted red line).
}
\label{fig:asymmetric}
\end{figure}

A more physically-motivated asymmetric line shape function was derived by Li et al.~\cite{Li1985} for the analysis of multiphoton resonances in the NO $A\,^2\Sigma^+ - X\,^2\Pi\,(0,0)$ band.
Their closed-form line shape model accounted for effects of the spatial and temporal distributions of the light intensity.
Here, we reproduce their line shape as a function of $\delta_L=\nu-\nu_0$, the laser frequency shift from the zero-field line position
\begin{equation}
S(\delta_0,\Gamma,\delta_L)=\kappa\int\limits_{0}^{\delta_0}d\delta'\frac{K\left( \ln\left( \frac{\delta_0}{\delta'}\right) \right) }{\delta'}G\left(\Gamma,\delta'-\delta_L \right), 
\label{LiProfile1}
\end{equation}
where $\delta_0$ is the maximum ac-Stark shift induced at the peak intensity $I_0$. 
$K$ contains the dependence on the temporal profile, as well as the transverse (Gaussian beam profile) and longitudinal intensity (focused) distribution, parametrized in \cite{Li1985} as
\begin{align}
K(x)=&\left[0.6366/x+2.087 e^{x/2}\right. \nonumber\\
     &\left. -e^{-x/2}(1.087+0.90x+0.45x^2+0.3x^3)\right]^{-1}.
\label{LiProfile2}
\end{align}
$G$ is a Gaussian distribution with full width at half maximum (FWHM) $\Gamma$,
\begin{align}
G\left(\Gamma,\delta'-\delta_L \right)=&\frac{1}{\sqrt{\pi}}\frac{2\sqrt{\ln 2}}{\Gamma} \nonumber\\
     &\times\exp \left(-\left( \frac{2\sqrt{\ln 2}}{\Gamma}\right)^2 (\delta'-\delta_L)^2 \right), 
\label{LiProfile3} 
\end{align}
that accounts for other sources of line broadening, such as the spectral width of the laser or natural linewidth.
The parameter $\kappa=1.189$ is a normalization factor which ensures that
\begin{equation}
\int\limits_{-\infty}^{\infty}S(\delta_0,\Gamma,\delta_L)d\delta_L=1
\label{LiProfile4} 
\end{equation}
for any $\Gamma$ and $\delta_0$.
It appears that the spatial and temporal intensity distribution were also treated in the investigations of Huo et al. \cite{Huo1985} and Girard et al. \cite{Girard1983} , but the expressions were not explicitly given. 
When comparing different probe intensity recordings, the normalized profile given by Eq.~(\ref{LiProfile1}), should be multiplied by a factor that scales with light intensity as $\sim I_0^2$, with the exact form given in Ref.~\cite{Li1985}.
Note that the error function in the skewed Gaussian model of Eq.~(\ref{eq:skewed}) effectively captures the result of the integration in Eq.~(\ref{LiProfile1}).
However, the physical interpretation of the $f_c$ and $\xi$ parameters from the skewed Gaussian model is ambiguous, while the background, $B(f)$, is an \emph{ad hoc} addition.

Li et al.~\cite{Li1985} presented intuitive explanations of qualitative behavior of the line profile at two extreme cases: 1) of a perfectly collimated probe beam and 2) of a conically-focused beam.
In case 1) the laser intensity is spatially homogeneous, so that the temporal intensity distribution is the dominant effect.
Almost all contribution to the resonant excitation comes from the peak of the pulse, causing the line peak position to be shifted by almost $\delta_0$.
In case 2) the strongly inhomogeneous spatial intensity plays the dominant role, and molecules located at the focus, having the highest Stark shift $\delta_0$, have a smaller contribution relative to those from the entire interaction volume.
The majority of the excited molecules come from a region of low intensities outside the focus, therefore the integrated line profile is only slightly shifted from the field-free resonance.
Our experimental conditions lie in between these two cases, where a loose focus is implemented and a molecular beam, that overlaps to within a few Rayleigh ranges of the laser beam, is employed.

\begin{figure}[!t]
\includegraphics[width=0.45\textwidth]{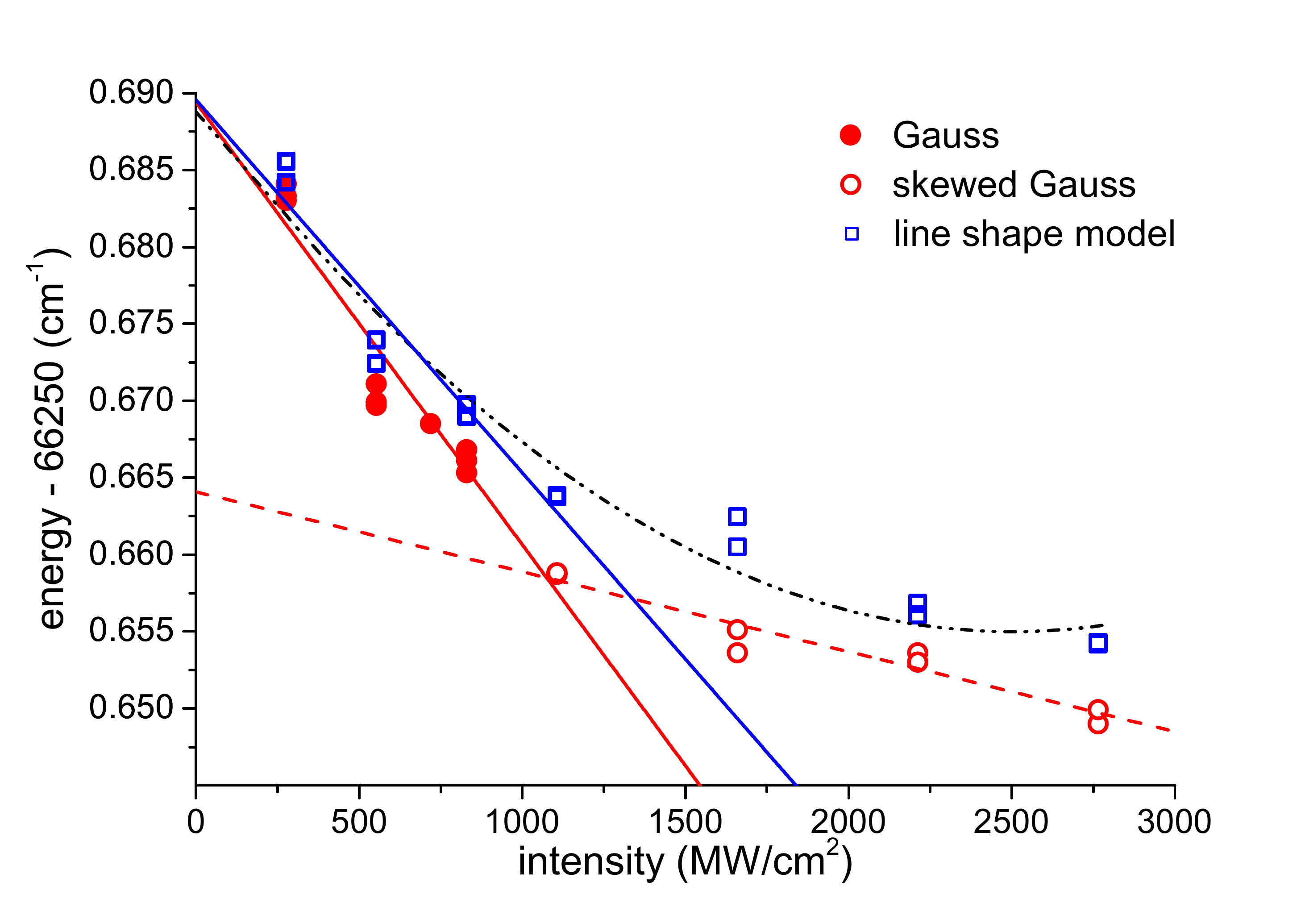}
\caption{Extracted line positions of the $F-X\,(0,11)\,Q(3)$ transition from measurements at different probe laser intensities.
Squares (blue) are obtained from fits using the line shape model of Eq.~(\ref{LiProfile1}), while circles (red) are obtained from Gaussian fits.
The solid circles (red) are symmetric profiles fitted with a simple Gaussian, while unfilled circles are fitted with skewed Gaussian profiles.
The solid lines are linear fits to low-intensity data points, while the dashed lines is a linear fit using the high-intensity points only.
The dash-dotted line is a second-order polynomial fit for the whole intensity range using the data points obtained from the line shape model (squares).
}
\label{fig:stark}
\end{figure}

Using the line shape model expressed in Eq.~(\ref{LiProfile1}) and appropriate experimental parameters, the asymmetric line profiles can be well fitted. 
For a recording at a particular intensity $I_0$, the maximum ac-Stark shift from the zero-field resonance, $\delta_0$, is obtained from the fit.
The line shape asymmetry, in particular the skew handedness, is consistent with the direction of the light shift observed at different intensities, validating the expected behavior from Eq.~(\ref{LiProfile1}).
In Fig.~\ref{fig:asymmetric}, fits using the physical line shape model and skewed Gaussian profile are shown for the $Q(3)$ transition recorded at $\sim9.4$ GW/cm$^2$.
The linear background $B(f)$ (dashed line) was necessary for a satisfactory fit with the skewed Gaussian, while no additional background functions were used for the line shape model.
The extracted line positions are indicated in Fig.~\ref{fig:asymmetric} by vertical lines above the profiles, where the difference in the line positions of the two fit functions amounts to about $0.006$ \wn.
For reference, the $f_c$ position obtained by using Eq.~(\ref{eq:skewed}) is also indicated by a dotted line, although this is not used further in the analysis.

A comparison of the $F-X$ (0,11) $Q(3)$ line positions extracted from the skewed Gaussian and from the line shape model based on Li et al.~\cite{Li1985} is shown in Fig.~\ref{fig:stark}.
The line positions show different trends that separate into low-intensity, with symmetric line profiles, and high-intensity subsets, with asymmetric ones.
The line profile models are in agreement at low intensities as expected, but show a discrepancy at higher intensities as explained above (see Fig.~\ref{fig:asymmetric}).
The extrapolated zero-intensity positions for the low-intensity measurements converge to within $0.0001$ \wn\ for both line profile models.

When using only the high-intensity data to extrapolate the zero-intensity frequency, a shift of $\sim0.02$ \wn\ with respect to the low-intensity subset is found.
The latter difference is a concern when only high-intensity data is available as in Ref.~\cite{Niu2015b} for the $F-X$ (3,12) band.
The ac-Stark coefficients, however, are an order of magnitude lower for the $F-X$ (3,12) band compared the present (0,11) band, thus any systematic offset is still expected to be within the uncertainty estimates in that study~\cite{Niu2015b}.
It is comforting to note that a second-order polynomial fit, also shown in Fig.~\ref{fig:stark} as dash-dotted curve, results in an extrapolated zero-field frequency that is within 0.0002 \wn\ of the low-intensity linear fits.

The ac-Stark shifts of the different $F-X$ (0,11) transitions exhibit a nonlinear dependence on intensity, contrary to the expected behavior in Eq.~(\ref{eq:Stark_coeff}).
A correlation is also observed between the nonlinearity and the ac-Stark coefficient $\alpha$, where the onset of nonlinearity occurs at a higher intensity for transitions with smaller $\alpha$.
The nonlinearity may indicate close proximity to a near-resonant state, which may signal the breakdown of the perturbative approximation in Eq.~(\ref{eq:deltaE}).
Possibly, this requires contributions beyond the second-order correction \cite{Girard1983} that lead to a higher-power dependence on intensity.
A similar behavior was observed by Liao and Bjorkholm in their study of the ac-Stark effect in two-photon excitation of sodium~\cite{Liao1975}.
In that investigation, they observed the nonlinear dependence of the ac-Stark shift at some probe detuning that is sufficiently close to resonance with an intermediate state. 

\begin{figure}[!t]
\includegraphics[width=0.45\textwidth]{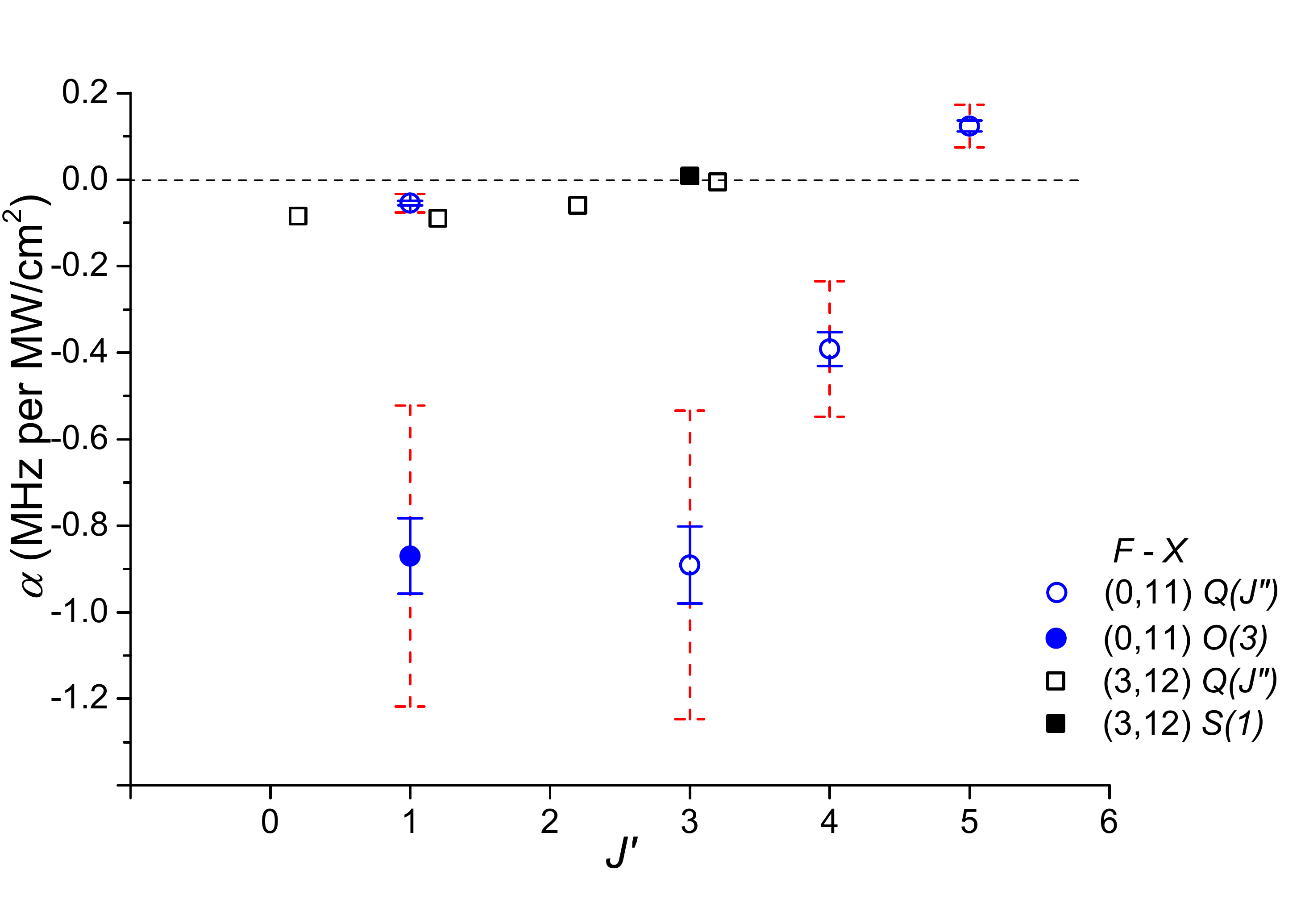}
\caption{
ac-Stark coefficients plotted against the upper $J'$-quantum number for the different transitions of the $F-X\,(0,11)$ band (blue circles).
The $\alpha$-values of the $F-X\,(3,12)$ transitions (black squares) from Ref.~\cite{Niu2015} are also plotted and displaced for clarity.
The datapoint indicated by a filled circle is obtained from the $F-X$ (0,11) $O(3)$ transition, while the filled rectangle is from the $F-X$ (3,12) $S(1)$ transition.
}
\label{fig:stark_coeff}
\end{figure}

\subsection{ac-Stark coefficients}

The ac-Stark coefficient $\alpha$, as defined in Eq.~(\ref{eq:Stark_coeff}), was obtained by Hannemann et al. for the \Hmol\ $EF{}^1\Sigma^+_g-X{}^1\Sigma^+_g$ (0,0) band, where they reported $+13(7)$ and $+6(4)$ MHz per MW/cm$^2$, respectively, for the $Q(0)$ and $Q(1)$ lines \cite{Hannemann2006}.
Investigations on the $E-X$ (0,1) band in Ref.~\cite{Dickenson2013}, and more extensively in Ref.~\cite{Niu2014}, also resulted in positive ac-Stark coefficients for $Q(J=0-3)$ transitions that are about an order of magnitude lower than for the $E-X$ (0,0) band.
Eyler and coworkers~\cite{Yiannopoulou2006} found that the ac-Stark slopes vary considerably, typically at a few tens of MHz per (MW/cm$^2$), for different transitions in the \Hmol\ $E-X$ system.
The Rhodes group in Chicago has performed a number of excitation studies with high-power lasers on the $EF-X$ system in hydrogen, where they also investigated optical Stark shifts (e.g.~\cite{Srinivasan1983b}).
Following excitation of molecular hydrogen by 193 nm radiation, intense stimulated emission on both the Lyman and Werner bands is observed.
Using excitations intensities of $\sim600$ GW/cm$^2$, they obtained shifts in the order of 2 MHz per MW/cm$^2$ for the $EF-X$ (2,0) band.
While the work of Vrakking et al.~\cite{Vrakking1993} was primarily on the detection sensitivity of REMPI on \Hmol\ $EF-X$ system, they also obtained ac-Stark coefficients of $\sim15$ MHz per MW/cm$^2$ similar to the value found in Ref.~\cite{Hannemann2006}.
Vrakking et al.~\cite{Vrakking1993} also made reference to a private communication with Hessler on the ac-Stark effect (shift) amounting to 3-6 MHz per MW/cm$^2$, presumably obtained in the study by Glab and Hessler~\cite{Glab1987}.

The ac-Stark coefficients $\alpha$ obtained in this study for the $F-X$ (0,11) transitions are plotted in Fig.~\ref{fig:stark_coeff}.
The error bars in the figure comprise two contributions, with the larger one dominated by the accuracy in the absolute determination of the probe intensity.
The difficulties include estimating the effective laser beam cross-sections in interaction volume, that should take into account the overlap of the counterpropagating probe beams and also the overlap of the photodissociation and ionization beams.
The smaller of the error bars are obtained from fits using relative intensities, shown here to emphasize that the differences in $\alpha$-coefficients are significant.
Also included in Fig.~\ref{fig:stark_coeff} are the Stark coefficients for $F-X$ (3,12) obtained in \cite{Niu2015}, that differ by about an order-of-magnitude with respect to the values obtained for $F-X$ (0,11).
The sign of the $F-X$ $\alpha$-coefficients are mostly negative for both bands except for the $F-X$ (0,11) $Q(5)$ line.

The different signs in the ac-Stark coefficients, which are positive for the $E-X$ transitions and mostly negative for the $F-X$ transitions, seem to be a feature of the ac-Stark effect in \Hmol.
For the $E-X$ (0,0) transitions probed at around 202 nm and $E-X$ (0,1) transitions probed at around 210 nm, all intermediates states are far off-resonant with respect to transitions from the $E$ or $X$ levels.
Using Eq.~(\ref{eq:deltaE}), the ac-Stark shift is estimated based on the approximation of Rhodes and coworkers~\cite{Pummer1983,Srinivasan1983a} for the $E-X$ (2,0) band.
In those studies, they assumed that the intermediate states are predominantly the Rydberg series at principal quantum number $n>2$, clustered at an average energy of $E_m = 14.7$ eV.
The present estimates of ac-Stark shifts via Eq.~(\ref{eq:deltaE}), using the probe frequency and intensity in Ref.~\cite{Hannemann2006}, are in agreement to within an order of magnitude of the $E-X$ (0,0) observations.
A similar estimate for $E-X$ (0,1) transitions are also within an order-of-magnitude agreement of the measurements in ~\cite{Dickenson2013,Niu2014}.
The blue-detuned probe in the aforementioned $E-X$ transitions explains the observed light shift direction as expected from Eq.~(\ref{eq:deltaE}).

\begin{table}
\centering
\caption{Measured two-photon transition energies of $F-X$ (0,11) band. All values in cm$^{-1}$.}
\label{tab:transitions}
\begin{tabular}{cc}
\toprule
Line & Experiment \\
\midrule
$Q(1)$ & 66\,438.2920\,(15) \\
$Q(3)$ & 66\,250.6874\,(15) \\
$Q(4)$ & 66\,105.8695\,(15) \\
$Q(5)$ & 65\,931.3315\,(15)\\
$O(3)$ & 66\,189.5591\,(15)  \\
\bottomrule
\end{tabular}
\end{table}

In $F-X$ excitation from $X,\,v=11$ and $v=12$, the fundamental probe wavelengths are around 300 nm.
Rydberg $np$ levels can be close to resonance at the probe wavelengths, a fact that we have exploited in the resonant ionization (using an additional laser source) for the REMPI detection.
This could explain the difference in sign of $\alpha$ between the $E-X$ and $F-X$ transitions.
Furthermore, Fig.~\ref{fig:stark_coeff} displays a trend the $F-X$ (0,11) transitions, where a small negative coefficient is observed for $Q(1)$, increasing in magnitude at $Q(3)$, decreasing in magnitude again and eventually becoming positive at $Q(5)$.
Interestingly, the $F-X$ (0,11) $O(3)$ transition displays a relatively large $\alpha$ with respect to that of the $Q(1)$ transition, despite having the same upper $F$ level $J'=1$.
The $Q(1)$ and $O(3)$ fundamental probe energies differ only by some 250 \wn, but this results in significant change in $\alpha$.
These phenomena hint towards a scenario where some near-resonant levels are accessed, so that the summation (and cancellation) of the contributions in Eq.~(\ref{eq:deltaE}) depend more sensitively on the detunings of the probe laser frequency with respect to $\nu_{mn}$.
The variation in $\alpha$-magnitudes as well as the change from negative to positive sign could be explained by a change from red- to blue-detuning when traversing across a dominant near-resonance intermediate level.
The latter behavior was expected in the NO investigation of Huo et al., where the $\alpha$-coefficients for individual $M$ sublevels varied from -4.1 to 53 MHz per MW/cm$^2$ \cite{Huo1985}.
The order of magnitude difference between $\alpha$-values of the $F-X$ (0,11) and (3,12) transitions may be due to less-favorable FCF overlaps of the near-resonant intermediate levels for the latter band.
This is correlated with the similar value of the $F-X$ (3,12) $Q(3)$ and $S(1)$ transition, displaying a different trend to that of the $Q(1)$ and $O(3)$ $\alpha$-coefficients in the $F-X$ (0,11) band.
The nonlinear intensity-dependence of the transitions discussed in the previous subsection, may be consistently explained by the same argument on the importance of near-resonant intermediate levels.
For a quantitative explanation of the $F-X$ ac-Stark shift, a more extensive theoretical study is necessary to account for the dense intermediate Rydberg levels involved that should also include considerations of FCF overlaps and dipole coupling strengths at the appropriate internuclear distance.  

\section{Results and Discussions}

The resulting two-photon transition energies for the $F\,^1\Sigma_g^+ - X\,^1\Sigma_g^+$ (0,11) lines are listed in Table~\ref{tab:transitions}.
The results presented are principally based on high-resolution measurements using transitions with symmetric linewidths narrower than 1 GHz.
This ensures that the ac-Stark shift extrapolation can be considered robust and reliable as discussed above.

Combination differences between appropriate transition pairs allow for the confirmation of transition assignments as well as consistency checks of the measurements, where most systematic uncertainty contributions cancel. The $Q(1)$ and $O(3)$ transitions share a common upper $EF$ level, and the energy difference of 248.7329(21) \wn\ gives the ground state splitting $X, v''=11, J''=1\rightarrow 3$. This can be compared to the theoretical splitting derived from Komasa et al.~\cite{Komasa2011} of 248.731(7) \wn.
In analogous fashion, the $Q(3)$ and $O(3)$ share the same lower $X$ level, which enables the extraction of the $EF, v'=1, J'=1\rightarrow 3$ energy splitting of 61.1194(21) \wn. This is in good agreement to the derived experimental splitting of 61.1191(10) \wn\ from Bailly et al.~\cite{Bailly2010}.

\begin{table}
\centering
\caption{Experimental and theoretical level energies of the $X\,^1\Sigma_g^+,v=11, J$ levels.}
\label{tab:levels}
\begin{tabular}{cccc}
\toprule
$J$& Experiment & Theory & Exp--Theo  \\
\midrule
1 & 32\,937.7554(16) & 32\,937.7494(53)	&0.0060(55)\\
3 & 33\,186.4791(16) & 33\,186.4802(52) &-0.0011(54)\\
4 & 33\,380.1025(33) & 33\,380.1015(52)	&0.0006(62)\\
5 & 33\,615.5371(18) & 33\,615.5293(51)	&0.0078(54)\\
\bottomrule
\end{tabular}
\end{table}

To extract the ground electronic $X\,^1\Sigma_g^+,\,v=11,J$ level energies, from the $EF-X$ transition energy measurements, we use the level energy values of the $F(v'=0)$ states determined by Bailly et al.~\cite{Bailly2010}.
The derived experimental level energies are listed in Table~\ref{tab:levels}, where the uncertainty is limited by the present $F-X$ determination except for the $J=4$. 
The calculated values obtained by Komasa et al.~\cite{Komasa2011} are also listed in Table~\ref{tab:levels}.
The experimental and theoretical values are in good agreement, except for $J=5$ that deviate by 1.5-$\sigma$.
The combined uncertainty of the difference is dominated by the theoretical uncertainty.
However, improvements in the calculations of the nonrelativistic energies, limited by the accuracy of fundamental constants $m_p/m_e$ and $R_\infty$, have recently been reported~\cite{Pachucki2016}, and improved calculations of QED corrections up to the $m\alpha^6$-order is anticipated.

As has been pointed out previously in Ref.~\cite{Niu2015b}, the uncertainties in the calculations~\cite{Komasa2011} are five times worse for the $v=8-11$ in comparison to the $v=0$ level energies.
Along with the previous measurements on the $X,\,v=12$ levels in Ref.~\cite{Niu2015b}, the measurements presented here probe the highest-uncertainty region of the most advanced first-principle quantum chemical calculations.

\section{Conclusion}

\Hmol\ transition energies of $F\,^1\Sigma^+_g(v'=0) - X\,^1\Sigma^+_g(v''=11)$ rovibrational states were determined at $0.0015$ cm$^{-1}$ absolute accuracies.
Enhanced detection efficiency was achieved by resonant excitation to autoionizing $7p\pi$ electronic Rydberg states, permitting excitation with low probe laser intensity that led to much narrower transitions due to reduced ac-Stark effects.
The asymmetric line broadening, induced by the ac-Stark effect, at high probe intensities was found to be well-explained by taking into account the spatial and temporal intensity beam profile of the probe laser.
The extracted ac-Stark coefficients for the different transitions $F-X$, as well as previously determined $E-X$ transitions, are consistent with qualitative expectations.
However, a quantitative explanation awaits detailed calculations of the ac-Stark effect that account for molecular structure, i.e. including a proper treatment of relevant intermediate states.

Using the $F$ level energies obtained by Bailly et al.~\cite{Bailly2010}, the level energies of $X\,(v=11, J=1,3-5)$ states are derived with accuracies better than 0.002 cm$^{-1}$ except for $J=3$, limited by $F$ level energy accuracy.
The derived experimental values are in excellent agreement with, thereby confirming, the results obtained from the most advanced and accurate molecular theory calculations.
The experimental binding energies reported here are about thrice more accurate than the present theoretical values, and may provide further stimulus towards advancements in the already impressive state-of-the-art \emph{ab initio} calculations.

\begin{acknowledgement}
This work is part of the research programme of the Foundation for Fundamental Research on Matter (FOM), which is part of the Netherlands Organisation for Scientific Research (NWO).
P. W. received support from LASERLAB-EUROPE within the EC's Seventh Framework Programme (Grant No. 284464).
W. U. received funding from the European Research Council (ERC) under the European Union’s Horizon 2020 research and innovation program (Grant No. 670168).
\end{acknowledgement}


\begin{thebibliography}{50}
\expandafter\ifx\csname natexlab\endcsname\relax\def\natexlab#1{#1}\fi
\expandafter\ifx\csname bibnamefont\endcsname\relax
  \def\bibnamefont#1{#1}\fi
\expandafter\ifx\csname bibfnamefont\endcsname\relax
  \def\bibfnamefont#1{#1}\fi
\expandafter\ifx\csname citenamefont\endcsname\relax
  \def\citenamefont#1{#1}\fi
\expandafter\ifx\csname url\endcsname\relax
  \def\url#1{\texttt{#1}}\fi
\expandafter\ifx\csname urlprefix\endcsname\relax\def\urlprefix{URL }\fi
\providecommand{\bibinfo}[2]{#2}
\providecommand{\eprint}[2][]{\url{#2}}

\bibitem[{\citenamefont{H\"{a}nsch}(1972)}]{Haensch1972}
\bibinfo{author}{\bibfnamefont{T.~W.} \bibnamefont{H\"{a}nsch}},
  \bibinfo{journal}{Appl. Opt.} \textbf{\bibinfo{volume}{11}},
  \bibinfo{pages}{895} (\bibinfo{year}{1972}).

\bibitem[{\citenamefont{H\"ansch et~al.}(1971)\citenamefont{H\"ansch, Levenson,
  and Schawlow}}]{Haensch1971}
\bibinfo{author}{\bibfnamefont{T.~W.} \bibnamefont{H\"ansch}},
  \bibinfo{author}{\bibfnamefont{M.~D.} \bibnamefont{Levenson}},
  \bibnamefont{and} \bibinfo{author}{\bibfnamefont{A.~L.}
  \bibnamefont{Schawlow}}, \bibinfo{journal}{Phys. Rev. Lett.}
  \textbf{\bibinfo{volume}{26}}, \bibinfo{pages}{946} (\bibinfo{year}{1971}).

\bibitem[{\citenamefont{H\"ansch et~al.}(1975)\citenamefont{H\"ansch, Lee,
  Wallenstein, and Wieman}}]{Haensch1975a}
\bibinfo{author}{\bibfnamefont{T.~W.} \bibnamefont{H\"ansch}},
  \bibinfo{author}{\bibfnamefont{S.~A.} \bibnamefont{Lee}},
  \bibinfo{author}{\bibfnamefont{R.}~\bibnamefont{Wallenstein}},
  \bibnamefont{and} \bibinfo{author}{\bibfnamefont{C.}~\bibnamefont{Wieman}},
  \bibinfo{journal}{Phys. Rev. Lett.} \textbf{\bibinfo{volume}{34}},
  \bibinfo{pages}{307} (\bibinfo{year}{1975}).

\bibitem[{\citenamefont{H\"ansch and Couillaud}(1980)}]{Haensch1980}
\bibinfo{author}{\bibfnamefont{T.~W.} \bibnamefont{H\"ansch}} \bibnamefont{and}
  \bibinfo{author}{\bibfnamefont{B.}~\bibnamefont{Couillaud}},
  \bibinfo{journal}{\oc} \textbf{\bibinfo{volume}{35}}, \bibinfo{pages}{441}
  (\bibinfo{year}{1980}).

\bibitem[{\citenamefont{Holzwarth et~al.}(2000)\citenamefont{Holzwarth, Udem,
  H\"ansch, Knight, Wadsworth, and Russell}}]{Holzwarth2000}
\bibinfo{author}{\bibfnamefont{R.}~\bibnamefont{Holzwarth}},
  \bibinfo{author}{\bibfnamefont{T.}~\bibnamefont{Udem}},
  \bibinfo{author}{\bibfnamefont{T.~W.} \bibnamefont{H\"ansch}},
  \bibinfo{author}{\bibfnamefont{J.~C.} \bibnamefont{Knight}},
  \bibinfo{author}{\bibfnamefont{W.~J.} \bibnamefont{Wadsworth}},
  \bibnamefont{and} \bibinfo{author}{\bibfnamefont{P.~S.~J.}
  \bibnamefont{Russell}}, \bibinfo{journal}{Phys. Rev. Lett.}
  \textbf{\bibinfo{volume}{85}}, \bibinfo{pages}{2264} (\bibinfo{year}{2000}).

\bibitem[{\citenamefont{H\"ansch}(2006)}]{Haensch2006}
\bibinfo{author}{\bibfnamefont{T.~W.} \bibnamefont{H\"ansch}},
  \bibinfo{journal}{Rev. Mod. Phys.} \textbf{\bibinfo{volume}{78}},
  \bibinfo{pages}{1297} (\bibinfo{year}{2006}).

\bibitem[{\citenamefont{Parthey et~al.}(2011)\citenamefont{Parthey, Matveev,
  Alnis, Bernhardt, Beyer, Holzwarth, Maistrou, Pohl, Predehl, Udem
  et~al.}}]{Parthey2011}
\bibinfo{author}{\bibfnamefont{C.~G.} \bibnamefont{Parthey}},
  \bibinfo{author}{\bibfnamefont{A.}~\bibnamefont{Matveev}},
  \bibinfo{author}{\bibfnamefont{J.}~\bibnamefont{Alnis}},
  \bibinfo{author}{\bibfnamefont{B.}~\bibnamefont{Bernhardt}},
  \bibinfo{author}{\bibfnamefont{A.}~\bibnamefont{Beyer}},
  \bibinfo{author}{\bibfnamefont{R.}~\bibnamefont{Holzwarth}},
  \bibinfo{author}{\bibfnamefont{A.}~\bibnamefont{Maistrou}},
  \bibinfo{author}{\bibfnamefont{R.}~\bibnamefont{Pohl}},
  \bibinfo{author}{\bibfnamefont{K.}~\bibnamefont{Predehl}},
  \bibinfo{author}{\bibfnamefont{T.}~\bibnamefont{Udem}}, \bibnamefont{et~al.},
  \bibinfo{journal}{Phys. Rev. Lett.} \textbf{\bibinfo{volume}{107}},
  \bibinfo{pages}{203001} (\bibinfo{year}{2011}).

\bibitem[{\citenamefont{Antognini et~al.}(2013)\citenamefont{Antognini, Nez,
  Schuhmann, Amaro, Biraben, Cardoso, Covita, Dax, Dhawan, Diepold
  et~al.}}]{Antognini2013}
\bibinfo{author}{\bibfnamefont{A.}~\bibnamefont{Antognini}},
  \bibinfo{author}{\bibfnamefont{F.}~\bibnamefont{Nez}},
  \bibinfo{author}{\bibfnamefont{K.}~\bibnamefont{Schuhmann}},
  \bibinfo{author}{\bibfnamefont{F.~D.} \bibnamefont{Amaro}},
  \bibinfo{author}{\bibfnamefont{F.}~\bibnamefont{Biraben}},
  \bibinfo{author}{\bibfnamefont{J.~M.~R.} \bibnamefont{Cardoso}},
  \bibinfo{author}{\bibfnamefont{D.~S.} \bibnamefont{Covita}},
  \bibinfo{author}{\bibfnamefont{A.}~\bibnamefont{Dax}},
  \bibinfo{author}{\bibfnamefont{S.}~\bibnamefont{Dhawan}},
  \bibinfo{author}{\bibfnamefont{M.}~\bibnamefont{Diepold}},
  \bibnamefont{et~al.}, \bibinfo{journal}{Science}
  \textbf{\bibinfo{volume}{339}}, \bibinfo{pages}{417} (\bibinfo{year}{2013}).

\bibitem[{\citenamefont{Komasa et~al.}(2011)\citenamefont{Komasa,
  Piszczatowski, \L{}ach, Przybytek, Jeziorski, and Pachucki}}]{Komasa2011}
\bibinfo{author}{\bibfnamefont{J.}~\bibnamefont{Komasa}},
  \bibinfo{author}{\bibfnamefont{K.}~\bibnamefont{Piszczatowski}},
  \bibinfo{author}{\bibfnamefont{G.}~\bibnamefont{\L{}ach}},
  \bibinfo{author}{\bibfnamefont{M.}~\bibnamefont{Przybytek}},
  \bibinfo{author}{\bibfnamefont{B.}~\bibnamefont{Jeziorski}},
  \bibnamefont{and} \bibinfo{author}{\bibfnamefont{K.}~\bibnamefont{Pachucki}},
  \bibinfo{journal}{J. Chem. Theory Comput.} \textbf{\bibinfo{volume}{7}},
  \bibinfo{pages}{3105} (\bibinfo{year}{2011}).

\bibitem[{\citenamefont{Pachucki and Komasa}(2014)}]{Pachucki2014}
\bibinfo{author}{\bibfnamefont{K.}~\bibnamefont{Pachucki}} \bibnamefont{and}
  \bibinfo{author}{\bibfnamefont{J.}~\bibnamefont{Komasa}},
  \bibinfo{journal}{J. Chem. Phys.} \textbf{\bibinfo{volume}{141}},
  \bibinfo{pages}{224103} (\bibinfo{year}{2014}).

\bibitem[{\citenamefont{Pachucki and Komasa}(2015)}]{Pachucki2015}
\bibinfo{author}{\bibfnamefont{K.}~\bibnamefont{Pachucki}} \bibnamefont{and}
  \bibinfo{author}{\bibfnamefont{J.}~\bibnamefont{Komasa}},
  \bibinfo{journal}{\jcp} \textbf{\bibinfo{volume}{143}},
  \bibinfo{pages}{034111} (\bibinfo{year}{2015}).

\bibitem[{\citenamefont{Korobov et~al.}(2014)\citenamefont{Korobov, Hilico, and
  Karr}}]{Korobov2014}
\bibinfo{author}{\bibfnamefont{V.~I.} \bibnamefont{Korobov}},
  \bibinfo{author}{\bibfnamefont{L.}~\bibnamefont{Hilico}}, \bibnamefont{and}
  \bibinfo{author}{\bibfnamefont{J.-P.} \bibnamefont{Karr}},
  \bibinfo{journal}{Phys. Rev. A} \textbf{\bibinfo{volume}{89}},
  \bibinfo{pages}{032511} (\bibinfo{year}{2014}).

\bibitem[{\citenamefont{Biesheuvel et~al.}(2016)\citenamefont{Biesheuvel, Karr,
  Hilico, Eikema, Ubachs, and Koelemeij}}]{Biesheuvel2016}
\bibinfo{author}{\bibfnamefont{J.}~\bibnamefont{Biesheuvel}},
  \bibinfo{author}{\bibfnamefont{J.-P.} \bibnamefont{Karr}},
  \bibinfo{author}{\bibfnamefont{L.}~\bibnamefont{Hilico}},
  \bibinfo{author}{\bibfnamefont{K.~S.~E.} \bibnamefont{Eikema}},
  \bibinfo{author}{\bibfnamefont{W.}~\bibnamefont{Ubachs}}, \bibnamefont{and}
  \bibinfo{author}{\bibfnamefont{J.~C.~J.} \bibnamefont{Koelemeij}},
  \bibinfo{journal}{Nature Comm.} \textbf{\bibinfo{volume}{7}},
  \bibinfo{pages}{10385} (\bibinfo{year}{2016}).

\bibitem[{\citenamefont{Karr et~al.}(2016)\citenamefont{Karr, Hilico,
  Koelemeij, and Korobov}}]{Karr2016}
\bibinfo{author}{\bibfnamefont{J.-P.} \bibnamefont{Karr}},
  \bibinfo{author}{\bibfnamefont{L.}~\bibnamefont{Hilico}},
  \bibinfo{author}{\bibfnamefont{J.}~\bibnamefont{Koelemeij}},
  \bibnamefont{and} \bibinfo{author}{\bibfnamefont{V.}~\bibnamefont{Korobov}}
  (\bibinfo{year}{2016}), \eprint{arXiv:1605.05456 [physics.atom-ph]}.

\bibitem[{\citenamefont{Pachucki and Komasa}(2016)}]{Pachucki2016}
\bibinfo{author}{\bibfnamefont{K.}~\bibnamefont{Pachucki}} \bibnamefont{and}
  \bibinfo{author}{\bibfnamefont{J.}~\bibnamefont{Komasa}},
  \bibinfo{journal}{\jcp} \textbf{\bibinfo{volume}{144}}, \bibinfo{eid}{164306}
  (\bibinfo{year}{2016}).

\bibitem[{\citenamefont{Ubachs et~al.}(2016)\citenamefont{Ubachs, Koelemeij,
  Eikema, and Salumbides}}]{Ubachs2016}
\bibinfo{author}{\bibfnamefont{W.}~\bibnamefont{Ubachs}},
  \bibinfo{author}{\bibfnamefont{J.}~\bibnamefont{Koelemeij}},
  \bibinfo{author}{\bibfnamefont{K.}~\bibnamefont{Eikema}}, \bibnamefont{and}
  \bibinfo{author}{\bibfnamefont{E.}~\bibnamefont{Salumbides}},
  \bibinfo{journal}{J. Mol. Spectros.} \textbf{\bibinfo{volume}{320}},
  \bibinfo{pages}{1} (\bibinfo{year}{2016}).

\bibitem[{\citenamefont{Liu et~al.}(2009)\citenamefont{Liu, Salumbides,
  Hollenstein, Koelemeij, Eikema, Ubachs, and Merkt}}]{Liu2009}
\bibinfo{author}{\bibfnamefont{J.}~\bibnamefont{Liu}},
  \bibinfo{author}{\bibfnamefont{E.~J.} \bibnamefont{Salumbides}},
  \bibinfo{author}{\bibfnamefont{U.}~\bibnamefont{Hollenstein}},
  \bibinfo{author}{\bibfnamefont{J.~C.~J.} \bibnamefont{Koelemeij}},
  \bibinfo{author}{\bibfnamefont{K.~S.~E.} \bibnamefont{Eikema}},
  \bibinfo{author}{\bibfnamefont{W.}~\bibnamefont{Ubachs}}, \bibnamefont{and}
  \bibinfo{author}{\bibfnamefont{F.}~\bibnamefont{Merkt}},
  \bibinfo{journal}{\jcp} \textbf{\bibinfo{volume}{130}},
  \bibinfo{pages}{174306} (\bibinfo{year}{2009}).

\bibitem[{\citenamefont{Salumbides et~al.}(2011)\citenamefont{Salumbides,
  Dickenson, Ivanov, and Ubachs}}]{Salumbides2011}
\bibinfo{author}{\bibfnamefont{E.~J.} \bibnamefont{Salumbides}},
  \bibinfo{author}{\bibfnamefont{G.~D.} \bibnamefont{Dickenson}},
  \bibinfo{author}{\bibfnamefont{T.~I.} \bibnamefont{Ivanov}},
  \bibnamefont{and} \bibinfo{author}{\bibfnamefont{W.}~\bibnamefont{Ubachs}},
  \bibinfo{journal}{\prl} \textbf{\bibinfo{volume}{107}},
  \bibinfo{pages}{043005} (\bibinfo{year}{2011}).

\bibitem[{\citenamefont{Dickenson et~al.}(2013)\citenamefont{Dickenson, Niu,
  Salumbides, Komasa, Eikema, Pachucki, and Ubachs}}]{Dickenson2013}
\bibinfo{author}{\bibfnamefont{G.~D.} \bibnamefont{Dickenson}},
  \bibinfo{author}{\bibfnamefont{M.~L.} \bibnamefont{Niu}},
  \bibinfo{author}{\bibfnamefont{E.~J.} \bibnamefont{Salumbides}},
  \bibinfo{author}{\bibfnamefont{J.}~\bibnamefont{Komasa}},
  \bibinfo{author}{\bibfnamefont{K.~S.~E.} \bibnamefont{Eikema}},
  \bibinfo{author}{\bibfnamefont{K.}~\bibnamefont{Pachucki}}, \bibnamefont{and}
  \bibinfo{author}{\bibfnamefont{W.}~\bibnamefont{Ubachs}},
  \bibinfo{journal}{\prl} \textbf{\bibinfo{volume}{110}},
  \bibinfo{pages}{193601} (\bibinfo{year}{2013}).

\bibitem[{\citenamefont{Salumbides et~al.}(2013)\citenamefont{Salumbides,
  Koelemeij, Komasa, Pachucki, Eikema, and Ubachs}}]{Salumbides2013}
\bibinfo{author}{\bibfnamefont{E.~J.} \bibnamefont{Salumbides}},
  \bibinfo{author}{\bibfnamefont{J.~C.~J.} \bibnamefont{Koelemeij}},
  \bibinfo{author}{\bibfnamefont{J.}~\bibnamefont{Komasa}},
  \bibinfo{author}{\bibfnamefont{K.}~\bibnamefont{Pachucki}},
  \bibinfo{author}{\bibfnamefont{K.~S.~E.} \bibnamefont{Eikema}},
  \bibnamefont{and} \bibinfo{author}{\bibfnamefont{W.}~\bibnamefont{Ubachs}},
  \bibinfo{journal}{\prd} \textbf{\bibinfo{volume}{87}},
  \bibinfo{pages}{112008} (\bibinfo{year}{2013}).

\bibitem[{\citenamefont{Salumbides et~al.}(2015)\citenamefont{Salumbides,
  Schellekens, Gato-Rivera, and Ubachs}}]{Salumbides2015b}
\bibinfo{author}{\bibfnamefont{E.~J.} \bibnamefont{Salumbides}},
  \bibinfo{author}{\bibfnamefont{A.~N.} \bibnamefont{Schellekens}},
  \bibinfo{author}{\bibfnamefont{B.}~\bibnamefont{Gato-Rivera}},
  \bibnamefont{and} \bibinfo{author}{\bibfnamefont{W.}~\bibnamefont{Ubachs}},
  \bibinfo{journal}{New J. Phys.} \textbf{\bibinfo{volume}{17}},
  \bibinfo{pages}{033015} (\bibinfo{year}{2015}).

\bibitem[{\citenamefont{Niu et~al.}(2015{\natexlab{a}})\citenamefont{Niu,
  Salumbides, and Ubachs}}]{Niu2015b}
\bibinfo{author}{\bibfnamefont{M.~L.} \bibnamefont{Niu}},
  \bibinfo{author}{\bibfnamefont{E.~J.} \bibnamefont{Salumbides}},
  \bibnamefont{and} \bibinfo{author}{\bibfnamefont{W.}~\bibnamefont{Ubachs}},
  \bibinfo{journal}{\jcp} \textbf{\bibinfo{volume}{143}},
  \bibinfo{pages}{081102} (\bibinfo{year}{2015}{\natexlab{a}}).

\bibitem[{\citenamefont{Steadman and Baer}(1989)}]{Steadman1989}
\bibinfo{author}{\bibfnamefont{J.}~\bibnamefont{Steadman}} \bibnamefont{and}
  \bibinfo{author}{\bibfnamefont{T.}~\bibnamefont{Baer}},
  \bibinfo{journal}{\jcp} \textbf{\bibinfo{volume}{91}}, \bibinfo{pages}{6113}
  (\bibinfo{year}{1989}).

\bibitem[{\citenamefont{Hannemann et~al.}(2007)\citenamefont{Hannemann,
  Salumbides, and Ubachs}}]{Hannemann2007}
\bibinfo{author}{\bibfnamefont{S.}~\bibnamefont{Hannemann}},
  \bibinfo{author}{\bibfnamefont{E.~J.} \bibnamefont{Salumbides}},
  \bibnamefont{and} \bibinfo{author}{\bibfnamefont{W.}~\bibnamefont{Ubachs}},
  \bibinfo{journal}{Opt. Lett.} \textbf{\bibinfo{volume}{32}},
  \bibinfo{pages}{1381} (\bibinfo{year}{2007}).

\bibitem[{\citenamefont{Fantz and W\"underlich}(2006)}]{Fantz2006}
\bibinfo{author}{\bibfnamefont{U.}~\bibnamefont{Fantz}} \bibnamefont{and}
  \bibinfo{author}{\bibfnamefont{D.}~\bibnamefont{W\"underlich}},
  \bibinfo{journal}{At. Data. Nucl. Data Tables} \textbf{\bibinfo{volume}{92}},
  \bibinfo{pages}{853 } (\bibinfo{year}{2006}).

\bibitem[{\citenamefont{Fantz and W\"underlich}(2004)}]{Fantz2004}
\bibinfo{author}{\bibfnamefont{U.}~\bibnamefont{Fantz}} \bibnamefont{and}
  \bibinfo{author}{\bibfnamefont{D.}~\bibnamefont{W\"underlich}}
  (\bibinfo{year}{2004}), \bibinfo{note}{monograph INDC(NDS)-457, University
  Augsburg, Germany}.

\bibitem[{\citenamefont{Glass-Maujean
  et~al.}(2013{\natexlab{a}})\citenamefont{Glass-Maujean, Jungen, Schmoranzer,
  Tulin, Knie, Reiss, and Ehresmann}}]{Glass-Maujean2013b}
\bibinfo{author}{\bibfnamefont{M.}~\bibnamefont{Glass-Maujean}},
  \bibinfo{author}{\bibfnamefont{C.}~\bibnamefont{Jungen}},
  \bibinfo{author}{\bibfnamefont{H.}~\bibnamefont{Schmoranzer}},
  \bibinfo{author}{\bibfnamefont{I.}~\bibnamefont{Tulin}},
  \bibinfo{author}{\bibfnamefont{A.}~\bibnamefont{Knie}},
  \bibinfo{author}{\bibfnamefont{P.}~\bibnamefont{Reiss}}, \bibnamefont{and}
  \bibinfo{author}{\bibfnamefont{A.}~\bibnamefont{Ehresmann}},
  \bibinfo{journal}{\jms} \textbf{\bibinfo{volume}{293-294}},
  \bibinfo{pages}{11} (\bibinfo{year}{2013}{\natexlab{a}}).

\bibitem[{\citenamefont{Glass-Maujean
  et~al.}(2013{\natexlab{b}})\citenamefont{Glass-Maujean, Jungen, Spielfiedel,
  Schmoranzer, Tulin, Knie, Reiss, and Ehresmann}}]{Glass-Maujean2013a}
\bibinfo{author}{\bibfnamefont{M.}~\bibnamefont{Glass-Maujean}},
  \bibinfo{author}{\bibfnamefont{C.}~\bibnamefont{Jungen}},
  \bibinfo{author}{\bibfnamefont{A.}~\bibnamefont{Spielfiedel}},
  \bibinfo{author}{\bibfnamefont{H.}~\bibnamefont{Schmoranzer}},
  \bibinfo{author}{\bibfnamefont{I.}~\bibnamefont{Tulin}},
  \bibinfo{author}{\bibfnamefont{A.}~\bibnamefont{Knie}},
  \bibinfo{author}{\bibfnamefont{P.}~\bibnamefont{Reiss}}, \bibnamefont{and}
  \bibinfo{author}{\bibfnamefont{A.}~\bibnamefont{Ehresmann}},
  \bibinfo{journal}{\jms} \textbf{\bibinfo{volume}{293-294}},
  \bibinfo{pages}{1} (\bibinfo{year}{2013}{\natexlab{b}}).

\bibitem[{\citenamefont{Glass-Maujean
  et~al.}(2013{\natexlab{c}})\citenamefont{Glass-Maujean, Jungen, Schmoranzer,
  Tulin, Knie, Reiss, and Ehresmann}}]{Glass-Maujean2013c}
\bibinfo{author}{\bibfnamefont{M.}~\bibnamefont{Glass-Maujean}},
  \bibinfo{author}{\bibfnamefont{C.}~\bibnamefont{Jungen}},
  \bibinfo{author}{\bibfnamefont{H.}~\bibnamefont{Schmoranzer}},
  \bibinfo{author}{\bibfnamefont{I.}~\bibnamefont{Tulin}},
  \bibinfo{author}{\bibfnamefont{A.}~\bibnamefont{Knie}},
  \bibinfo{author}{\bibfnamefont{P.}~\bibnamefont{Reiss}}, \bibnamefont{and}
  \bibinfo{author}{\bibfnamefont{A.}~\bibnamefont{Ehresmann}},
  \bibinfo{journal}{\jms} \textbf{\bibinfo{volume}{293-294}},
  \bibinfo{pages}{19} (\bibinfo{year}{2013}{\natexlab{c}}).

\bibitem[{\citenamefont{Herzberg and Jungen}(1972)}]{Herzberg1972}
\bibinfo{author}{\bibfnamefont{G.}~\bibnamefont{Herzberg}} \bibnamefont{and}
  \bibinfo{author}{\bibfnamefont{C.}~\bibnamefont{Jungen}},
  \bibinfo{journal}{J. Mol. Spectrosc.} \textbf{\bibinfo{volume}{41}},
  \bibinfo{pages}{425} (\bibinfo{year}{1972}).

\bibitem[{\citenamefont{Glass-Maujean et~al.}(2010)\citenamefont{Glass-Maujean,
  Jungen, Schmoranzer, Knie, Haar, Hentges, Kielich, J\"{a}nk\"{a}l\"{a}, and
  Ehresmann}}]{Glass-Maujean2010a}
\bibinfo{author}{\bibfnamefont{M.}~\bibnamefont{Glass-Maujean}},
  \bibinfo{author}{\bibfnamefont{C.}~\bibnamefont{Jungen}},
  \bibinfo{author}{\bibfnamefont{H.}~\bibnamefont{Schmoranzer}},
  \bibinfo{author}{\bibfnamefont{A.}~\bibnamefont{Knie}},
  \bibinfo{author}{\bibfnamefont{I.}~\bibnamefont{Haar}},
  \bibinfo{author}{\bibfnamefont{R.}~\bibnamefont{Hentges}},
  \bibinfo{author}{\bibfnamefont{W.}~\bibnamefont{Kielich}},
  \bibinfo{author}{\bibfnamefont{K.}~\bibnamefont{J\"{a}nk\"{a}l\"{a}}},
  \bibnamefont{and}
  \bibinfo{author}{\bibfnamefont{A.}~\bibnamefont{Ehresmann}},
  \bibinfo{journal}{Phys. Rev. Lett.} \textbf{\bibinfo{volume}{104}},
  \bibinfo{pages}{183002} (\bibinfo{year}{2010}).

\bibitem[{\citenamefont{Xu et~al.}(2000)\citenamefont{Xu, van Dierendonck,
  Hogervorst, and Ubachs}}]{Xu2000}
\bibinfo{author}{\bibfnamefont{S.}~\bibnamefont{Xu}},
  \bibinfo{author}{\bibfnamefont{R.}~\bibnamefont{van Dierendonck}},
  \bibinfo{author}{\bibfnamefont{W.}~\bibnamefont{Hogervorst}},
  \bibnamefont{and} \bibinfo{author}{\bibfnamefont{W.}~\bibnamefont{Ubachs}},
  \bibinfo{journal}{J. Mol. Spectr.} \textbf{\bibinfo{volume}{201}},
  \bibinfo{pages}{256} (\bibinfo{year}{2000}).

\bibitem[{\citenamefont{Bodermann et~al.}(2002)\citenamefont{Bodermann,
  Kn\"{o}ckel, and Tiemann}}]{Bodermann2002}
\bibinfo{author}{\bibfnamefont{B.}~\bibnamefont{Bodermann}},
  \bibinfo{author}{\bibfnamefont{H.}~\bibnamefont{Kn\"{o}ckel}},
  \bibnamefont{and} \bibinfo{author}{\bibfnamefont{E.}~\bibnamefont{Tiemann}},
  \bibinfo{journal}{Eur. J. Phys. D} \textbf{\bibinfo{volume}{19}},
  \bibinfo{pages}{31} (\bibinfo{year}{2002}).

\bibitem[{\citenamefont{Niu et~al.}(2015{\natexlab{b}})\citenamefont{Niu,
  Ramirez, Salumbides, and Ubachs}}]{Niu2015}
\bibinfo{author}{\bibfnamefont{M.~L.} \bibnamefont{Niu}},
  \bibinfo{author}{\bibfnamefont{F.}~\bibnamefont{Ramirez}},
  \bibinfo{author}{\bibfnamefont{E.~J.} \bibnamefont{Salumbides}},
  \bibnamefont{and} \bibinfo{author}{\bibfnamefont{W.}~\bibnamefont{Ubachs}},
  \bibinfo{journal}{\jcp} \textbf{\bibinfo{volume}{142}},
  \bibinfo{pages}{044302} (\bibinfo{year}{2015}{\natexlab{b}}).

\bibitem[{\citenamefont{Bakos}(1977)}]{Bakos1977}
\bibinfo{author}{\bibfnamefont{J.}~\bibnamefont{Bakos}},
  \bibinfo{journal}{Phys. Rep.} \textbf{\bibinfo{volume}{31}},
  \bibinfo{pages}{209 } (\bibinfo{year}{1977}).

\bibitem[{\citenamefont{Ye et~al.}(2008)\citenamefont{Ye, Kimble, and
  Katori}}]{Ye2008}
\bibinfo{author}{\bibfnamefont{J.}~\bibnamefont{Ye}},
  \bibinfo{author}{\bibfnamefont{H.~J.} \bibnamefont{Kimble}},
  \bibnamefont{and} \bibinfo{author}{\bibfnamefont{H.}~\bibnamefont{Katori}},
  \bibinfo{journal}{Science} \textbf{\bibinfo{volume}{320}},
  \bibinfo{pages}{1734} (\bibinfo{year}{2008}).

\bibitem[{\citenamefont{Otis and Johnson}(1981)}]{Otis1981}
\bibinfo{author}{\bibfnamefont{C.~E.} \bibnamefont{Otis}} \bibnamefont{and}
  \bibinfo{author}{\bibfnamefont{P.~M.} \bibnamefont{Johnson}},
  \bibinfo{journal}{Chem. Phys. Lett.} \textbf{\bibinfo{volume}{83}},
  \bibinfo{pages}{73 } (\bibinfo{year}{1981}).

\bibitem[{\citenamefont{Huo et~al.}(1985)\citenamefont{Huo, Gross, and
  McKenzie}}]{Huo1985}
\bibinfo{author}{\bibfnamefont{W.~M.} \bibnamefont{Huo}},
  \bibinfo{author}{\bibfnamefont{K.~P.} \bibnamefont{Gross}}, \bibnamefont{and}
  \bibinfo{author}{\bibfnamefont{R.~L.} \bibnamefont{McKenzie}},
  \bibinfo{journal}{Phys. Rev. Lett.} \textbf{\bibinfo{volume}{54}},
  \bibinfo{pages}{1012} (\bibinfo{year}{1985}).

\bibitem[{\citenamefont{Girard et~al.}(1983)\citenamefont{Girard, Billy, Vigue,
  and Lehmann}}]{Girard1983}
\bibinfo{author}{\bibfnamefont{B.}~\bibnamefont{Girard}},
  \bibinfo{author}{\bibfnamefont{N.}~\bibnamefont{Billy}},
  \bibinfo{author}{\bibfnamefont{J.}~\bibnamefont{Vigue}}, \bibnamefont{and}
  \bibinfo{author}{\bibfnamefont{J.}~\bibnamefont{Lehmann}},
  \bibinfo{journal}{Chem. Phys. Lett.} \textbf{\bibinfo{volume}{102}},
  \bibinfo{pages}{168 } (\bibinfo{year}{1983}).

\bibitem[{\citenamefont{Srinivasan
  et~al.}(1983{\natexlab{a}})\citenamefont{Srinivasan, Egger, Luk, Pummer, and
  Rhodes}}]{Srinivasan1983b}
\bibinfo{author}{\bibfnamefont{T.}~\bibnamefont{Srinivasan}},
  \bibinfo{author}{\bibfnamefont{H.}~\bibnamefont{Egger}},
  \bibinfo{author}{\bibfnamefont{T.}~\bibnamefont{Luk}},
  \bibinfo{author}{\bibfnamefont{H.}~\bibnamefont{Pummer}}, \bibnamefont{and}
  \bibinfo{author}{\bibfnamefont{C.}~\bibnamefont{Rhodes}},
  \bibinfo{journal}{IEEE J. Quant. Electr.} \textbf{\bibinfo{volume}{19}},
  \bibinfo{pages}{1874} (\bibinfo{year}{1983}{\natexlab{a}}).

\bibitem[{\citenamefont{Vrakking et~al.}(1993)\citenamefont{Vrakking, Bracker,
  Suzuki, and Lee}}]{Vrakking1993}
\bibinfo{author}{\bibfnamefont{M.}~\bibnamefont{Vrakking}},
  \bibinfo{author}{\bibfnamefont{A.}~\bibnamefont{Bracker}},
  \bibinfo{author}{\bibfnamefont{T.}~\bibnamefont{Suzuki}}, \bibnamefont{and}
  \bibinfo{author}{\bibfnamefont{Y.}~\bibnamefont{Lee}},
  \bibinfo{journal}{\rsi} \textbf{\bibinfo{volume}{64}}, \bibinfo{pages}{3}
  (\bibinfo{year}{1993}).

\bibitem[{\citenamefont{Yiannopoulou et~al.}(2006)\citenamefont{Yiannopoulou,
  Melikechi, Gangopadhyay, Meiners, Cheng, and Eyler}}]{Yiannopoulou2006}
\bibinfo{author}{\bibfnamefont{A.}~\bibnamefont{Yiannopoulou}},
  \bibinfo{author}{\bibfnamefont{N.}~\bibnamefont{Melikechi}},
  \bibinfo{author}{\bibfnamefont{S.}~\bibnamefont{Gangopadhyay}},
  \bibinfo{author}{\bibfnamefont{J.~C.} \bibnamefont{Meiners}},
  \bibinfo{author}{\bibfnamefont{C.~H.} \bibnamefont{Cheng}}, \bibnamefont{and}
  \bibinfo{author}{\bibfnamefont{E.~E.} \bibnamefont{Eyler}},
  \bibinfo{journal}{\pra} \textbf{\bibinfo{volume}{73}},
  \bibinfo{pages}{022506} (\bibinfo{year}{2006}).

\bibitem[{\citenamefont{Hannemann et~al.}(2006)\citenamefont{Hannemann,
  Salumbides, Witte, Zinkstok, van Duijn, Eikema, and Ubachs}}]{Hannemann2006}
\bibinfo{author}{\bibfnamefont{S.}~\bibnamefont{Hannemann}},
  \bibinfo{author}{\bibfnamefont{E.~J.} \bibnamefont{Salumbides}},
  \bibinfo{author}{\bibfnamefont{S.}~\bibnamefont{Witte}},
  \bibinfo{author}{\bibfnamefont{R.~T.} \bibnamefont{Zinkstok}},
  \bibinfo{author}{\bibfnamefont{E.~J.} \bibnamefont{van Duijn}},
  \bibinfo{author}{\bibfnamefont{K.~S.~E.} \bibnamefont{Eikema}},
  \bibnamefont{and} \bibinfo{author}{\bibfnamefont{W.}~\bibnamefont{Ubachs}},
  \bibinfo{journal}{Phys. Rev. A} \textbf{\bibinfo{volume}{74}},
  \bibinfo{pages}{062514} (\bibinfo{year}{2006}).

\bibitem[{\citenamefont{Li et~al.}(1985)\citenamefont{Li, Yang, and
  Johnson}}]{Li1985}
\bibinfo{author}{\bibfnamefont{L.}~\bibnamefont{Li}},
  \bibinfo{author}{\bibfnamefont{B.-X.} \bibnamefont{Yang}}, \bibnamefont{and}
  \bibinfo{author}{\bibfnamefont{P.~M.} \bibnamefont{Johnson}},
  \bibinfo{journal}{J. Opt. Soc. Am. B} \textbf{\bibinfo{volume}{2}},
  \bibinfo{pages}{748} (\bibinfo{year}{1985}).

\bibitem[{\citenamefont{Liao and Bjorkholm}(1975)}]{Liao1975}
\bibinfo{author}{\bibfnamefont{P.~F.} \bibnamefont{Liao}} \bibnamefont{and}
  \bibinfo{author}{\bibfnamefont{J.~E.} \bibnamefont{Bjorkholm}},
  \bibinfo{journal}{Phys. Rev. Lett.} \textbf{\bibinfo{volume}{34}},
  \bibinfo{pages}{1} (\bibinfo{year}{1975}).

\bibitem[{\citenamefont{Niu et~al.}(2014)\citenamefont{Niu, Salumbides,
  Dickenson, Eikema, and Ubachs}}]{Niu2014}
\bibinfo{author}{\bibfnamefont{M.~L.} \bibnamefont{Niu}},
  \bibinfo{author}{\bibfnamefont{E.~J.} \bibnamefont{Salumbides}},
  \bibinfo{author}{\bibfnamefont{G.~D.} \bibnamefont{Dickenson}},
  \bibinfo{author}{\bibfnamefont{K.~S.~E.} \bibnamefont{Eikema}},
  \bibnamefont{and} \bibinfo{author}{\bibfnamefont{W.}~\bibnamefont{Ubachs}},
  \bibinfo{journal}{\jms} \textbf{\bibinfo{volume}{300}}, \bibinfo{pages}{44}
  (\bibinfo{year}{2014}).

\bibitem[{\citenamefont{Glab and Hessler}(1987)}]{Glab1987}
\bibinfo{author}{\bibfnamefont{W.~L.} \bibnamefont{Glab}} \bibnamefont{and}
  \bibinfo{author}{\bibfnamefont{J.~P.} \bibnamefont{Hessler}},
  \bibinfo{journal}{Phys. Rev. A} \textbf{\bibinfo{volume}{35}},
  \bibinfo{pages}{2102} (\bibinfo{year}{1987}).

\bibitem[{\citenamefont{Pummer et~al.}(1983)\citenamefont{Pummer, Egger, Luk,
  Srinivasan, and Rhodes}}]{Pummer1983}
\bibinfo{author}{\bibfnamefont{H.}~\bibnamefont{Pummer}},
  \bibinfo{author}{\bibfnamefont{H.}~\bibnamefont{Egger}},
  \bibinfo{author}{\bibfnamefont{T.~S.} \bibnamefont{Luk}},
  \bibinfo{author}{\bibfnamefont{T.}~\bibnamefont{Srinivasan}},
  \bibnamefont{and} \bibinfo{author}{\bibfnamefont{C.~K.}
  \bibnamefont{Rhodes}}, \bibinfo{journal}{Phys. Rev. A}
  \textbf{\bibinfo{volume}{28}}, \bibinfo{pages}{795} (\bibinfo{year}{1983}).

\bibitem[{\citenamefont{Srinivasan
  et~al.}(1983{\natexlab{b}})\citenamefont{Srinivasan, Egger, Pummer, and
  Rhodes}}]{Srinivasan1983a}
\bibinfo{author}{\bibfnamefont{T.}~\bibnamefont{Srinivasan}},
  \bibinfo{author}{\bibfnamefont{H.}~\bibnamefont{Egger}},
  \bibinfo{author}{\bibfnamefont{H.}~\bibnamefont{Pummer}}, \bibnamefont{and}
  \bibinfo{author}{\bibfnamefont{C.}~\bibnamefont{Rhodes}},
  \bibinfo{journal}{IEEE J. Quant. Electr.} \textbf{\bibinfo{volume}{19}},
  \bibinfo{pages}{1270} (\bibinfo{year}{1983}{\natexlab{b}}).

\bibitem[{\citenamefont{Bailly et~al.}(2010)\citenamefont{Bailly, Salumbides,
  Vervloet, and Ubachs}}]{Bailly2010}
\bibinfo{author}{\bibfnamefont{D.}~\bibnamefont{Bailly}},
  \bibinfo{author}{\bibfnamefont{E.}~\bibnamefont{Salumbides}},
  \bibinfo{author}{\bibfnamefont{M.}~\bibnamefont{Vervloet}}, \bibnamefont{and}
  \bibinfo{author}{\bibfnamefont{W.}~\bibnamefont{Ubachs}},
  \bibinfo{journal}{\molp} \textbf{\bibinfo{volume}{108}}, \bibinfo{pages}{827}
  (\bibinfo{year}{2010}).

\end{thebibliography}
\end{document}